\documentclass[journal,onecolumn,12pt,draftcls,onecolumn]{IEEEtran}
\usepackage{cite}
\usepackage{graphicx}
\usepackage{caption}
\ifCLASSOPTIONcompsoc
  \usepackage[caption=false,font=normalsize,labelfont=sf,textfont=sf]{subfig}
\else
  \usepackage[caption=false,font=footnotesize]{subfig}
\fi
\usepackage{amsmath}
\usepackage{amsthm}
\usepackage{algorithmic}
\usepackage{array}
\usepackage{tikz}
\usepackage{color}

\newtheorem{lemma}{Lemma}
\newtheorem{Proposition}{Proposition}
\newtheorem{corollary}{Corollary}
\newtheorem{Remark}{Remark}

\begin{document}
\title{A Novel Network NOMA Scheme for Downlink Coordinated Three-Point Systems}
\author{Yanshi Sun, Zhiguo Ding, \IEEEmembership{Senior Member, IEEE}, Xuchu Dai, and George K. Karagiannidis, \IEEEmembership{Fellow, IEEE}
\thanks{Y. Sun and X. Dai are with the Key Laboratory of Wireless-Optical Communications, Chinese Academy of Sciences, School of Information Science and Technology, University of Science and Technology of China,
No. 96 Jinzhai Road, Hefei, Anhui Province, 230026, P. R. China. (email: sys@mail.ustc.edu.cn, daixc@ustc.edu.cn).

Z. Ding is with the School of Computer and Communications, Lancaster University, LA1 4YW, U.K. (e-mail: z.ding@lancaster.ac.uk).

G. K. Karagiannidis is with the Provincial Key Laboratory of Information
Coding and Transmission, Southwest Jiaotong University, Sichuan 611756,
China, and also with the Electrical and Computer Engineering Department,
Aristotle University of Thessaloniki, Thessaloniki GR-54124, Greece. (email: geokarag@auth.gr).
}}
\maketitle
\vspace{-4em}
\begin{abstract}
To {\color{blue}study the feasibility of network non-orthogonal multiple access (N-NOMA) techniques, this paper proposes a N-NOMA scheme for a downlink coordinated multipoint (CoMP) communication scenario, with randomly deployed users}. In the considered {\color{black}N-NOMA} scheme, superposition coding (SC) is employed to serve cell-edge users {\color{black} as well as} users close to base stations (BSs)  simultaneously, and distributed analog beamforming by the BSs to meet the cell-edge user's quality of service (QoS) requirements. The combination of SC and distributed analog beamforming significantly complicates the expressions for the signal-to-interference-plus-noise ratio (SINR) at the reveiver, which makes the performance analysis particularly challenging. {\color{black}However, by using} rational approximations, insightful analytical results are obtained in order to characterize the outage performance of the considered {\color{black}N-NOMA} scheme. Computer simulation results are provided to show the superior performance of the proposed scheme {\color{black}as well as to} demonstrate the accuracy of the analytical results.
\end{abstract}
\IEEEpeerreviewmaketitle

\section{Introduction}
\IEEEPARstart{N}on-orthogonal multiple access (NOMA) has been recognized as a promising multiple access technique for the fifth generation (5G) mobile networks, due to its superior spectral efficiency\cite{saito2013non,ding2017application,kim2013non}. By using NOMA, multiple users can be served simultaneously at the same time, frequency, and spreading code. The key idea of NOMA is to apply superposition coding (SC) at the transmitter in order to efficiently mix multiple users' signals \cite{cover2012elements,choi2015minimum}. Furthermore, at the transmitter, NOMA power allocation is utilized by exploiting the difference among the users' channel conditions, i.e., users with poorer channel conditions are allocated more transmission power. On the other hand,  at the receiver side, users with better channel conditions apply successive interference cancellation (SIC) in order to separate the received mixture  \cite{higuchi2015non}.

The performance of NOMA with randomly deployed users has been studied in \cite{ding2014performance}, and the user fairness of NOMA was investigated in \cite{timotheou2015fairness}. In \cite{ding2014impact}, the impact of user pairing on NOMA has been studied, where the power allocation coefficients are chosen to meet the predefined users' quality of service (QoS) requirements. Furthermore, the design of uplink NOMA has been proposed and studied in \cite{al2014uplink}, while the application of multiple-input multiple-output (MIMO) technologies to NOMA has been studied in \cite{ding2016application,sun2015ergodic,ding2016general}. Recently, NOMA has been applied {\color{black}to} massive MIMO and millimeter wave (mmWave) networks \cite{ding2017noma}, {\color{black}as well as}  to visible light communication{\color{black}s} \cite{marshoud2016non}.

Network MIMO is a family of smart antenna techniques, where each user in a wireless system is served by multiple {\color{black}base stations (BSs)} or access points (APs), which are within the {\color{black}service} range of the user \cite{huang2009increasing,venkatesan2007network}. As a typical representative of network MIMO, coordinated multipoint (CoMP) has been recognized as an important enhancement for LTE-A\cite{jungnickel2009coordinated,sawahashi2010coordinated,irmer2011coordinated}. The main benefit to use CoMP is to improve the cell-edge users' data rates and hence improve the cell coverage. For CoMP transmission in downlink, various schemes can be applied, such as dynamic cell selection and coordinated beamforming \cite{venturino2010coordinated,dahrouj2010coordinated}. A drawback of these schemes is, that all the associated {\color{black}BSs} need to allocate the same channel resource block {\color{black}with the} cell-edge user. {\color{black}When} orthogonal multiple access (OMA) is used, this channel resource block cannot be accessed by other users, and hence, the spectral efficiency becomes worse as the number of cell-edge users increases. {\color{black}In order to} deal with this problem, {\color{black}network NOMA (N-NOMA)} has been proposed for CoMP \cite{choi2014non,tian2016performance}. In \cite{choi2014non}, NOMA with SC was employed in a CoMP system with two BSs to provide robust service to cell-edge users and to users close to BSs concurrently. Furthermore, in \cite{choi2014non}, Alamouti code, originally proposed in \cite{alamouti1998simple}, has been applied to improve the cell-edge user's reception reliability. In \cite{tian2016performance}, NOMA with  opportunistic BS or AP selection has been studied for CoMP. Both schemes in \cite{choi2014non} and \cite{tian2016performance} enlarges the system throughput, which demonstrates the superior performance of N-NOMA.

This paper {\color{black}proposes} a {\color{black}novel} N-NOMA scheme for a downlink CoMP system with randomly deployed users. {\color{blue}The users are divided into two categories: the cell-edge users and the near users. The cell-edge users are far from the BSs and are with strict QoS requirements, while the near users are close to the BSs and are served opportunistically.} The contributions of this paper are listed as {\color{black}follow.}
\begin{itemize}
  \item A downlink CoMP transmission system with three BSs and  randomly deployed users, is considered. A {\color{black}novel} N-NOMA scheme, {\color{black}which combines} SC with distributed analog beamforming is proposed. Specifically, SC is {\color{black}used} to support a cell-edge user and users close to the BSs simultaneously, and distributed analog beamforming {\color{black}in order to} improve the QoS of the cell-edge user.
      The reason for using distributed analog beamforming {\color{black}is} twofold.
      On the one hand, analog beamforming modifies the signal's phase only, which means the BS only needs to know {\color{black}the phase of the channel}. Thus, different BSs do not need to exchange instantaneous channel state information (CSI), {\color{black}and} as a result, system overhead is reduced.
      On the other hand, due to the knowledge of the channel phase information, distributed analog beamforming can utilize the spatial degrees of freedom more efficiently than the space time block code (STBC) scheme used in \cite{choi2014non}{\color{black}{\color{black},} without using any CSI}.
  \item Outage probability is used in this paper as the criterion to characterize the performance of the proposed scheme. The reason is that {\color{black}outage probability} not only bounds the error probability of detection tightly, but also can be used to {\color{black}efficiently evaluate} the outage sum rate/capacity.
      {\color{black}However,} there are two {\color{black}obstacles} to overcome in order to characterize the outage probability of the considered N-NOMA scheme, which makes the {\color{black}analysis} very challenging{\color{black}:
       a)} due to the combination of SC and distributed analog beamforming, the expression for the corresponding signal-to-interference-plus-noise ratio (SINR) is very different from those of the {\color{black}conventional} communication schemes, {\color{black}b)} how to capture the impact of the random {\color{black}users'} locations on the performance of the considered N-NOMA scheme. By using rational approximations and rigorous derivations, the above difficulties can be perfectly settled, and closed-form analytical results of the outage probabilities achieved by the cell-edge user and near users are obtained.
  \item To get more insight into the proposed scheme, the impact of the system parameters, such as user locations, distances between BSs and power allocation coefficients, {\color{black}on the performance of the proposed N-NOMA, is discussed}. For comparison purposes, a conventional OMA scheme, which uses distributed digital beamforming and serves {\color{black}only} the cell-edge user, is considered. {\color{black}This} comparison is facilitated, by using the analytical as well as {\color{black}the} computer simulation results.
\end{itemize}

{\color{black}The rest of this} paper is organized as follows. Section II illustrates the N-NOMA system model. Section III characterizes the outage performance of the proposed N-NOMA scheme. Section IV provides numerical results to demonstrate the performance of the proposed N-NOMA system and also verify the accuracy of the developed analytical results. Section V concludes the paper. Finally, Appendixes collect the proofs of the obtained analytical results.
\section{SYSTEM MODEL}
\begin{figure*}[!t]
\vspace{-1em}
\setlength{\belowcaptionskip}{-1em}   
\centering
\subfloat[]{\includegraphics[width=3.2in]{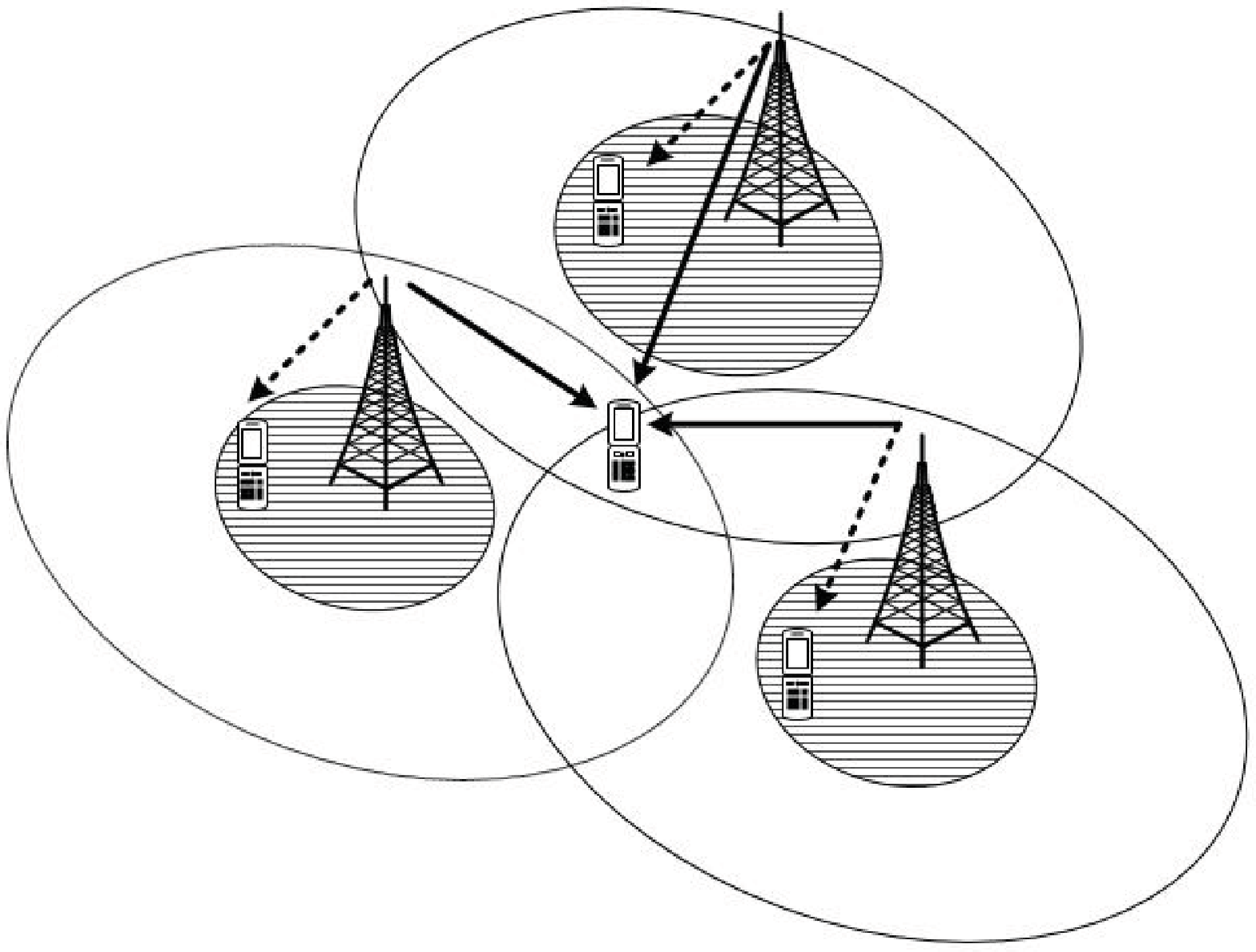}%
\label{system_a}}
\subfloat[]{\includegraphics[width=3.2in]{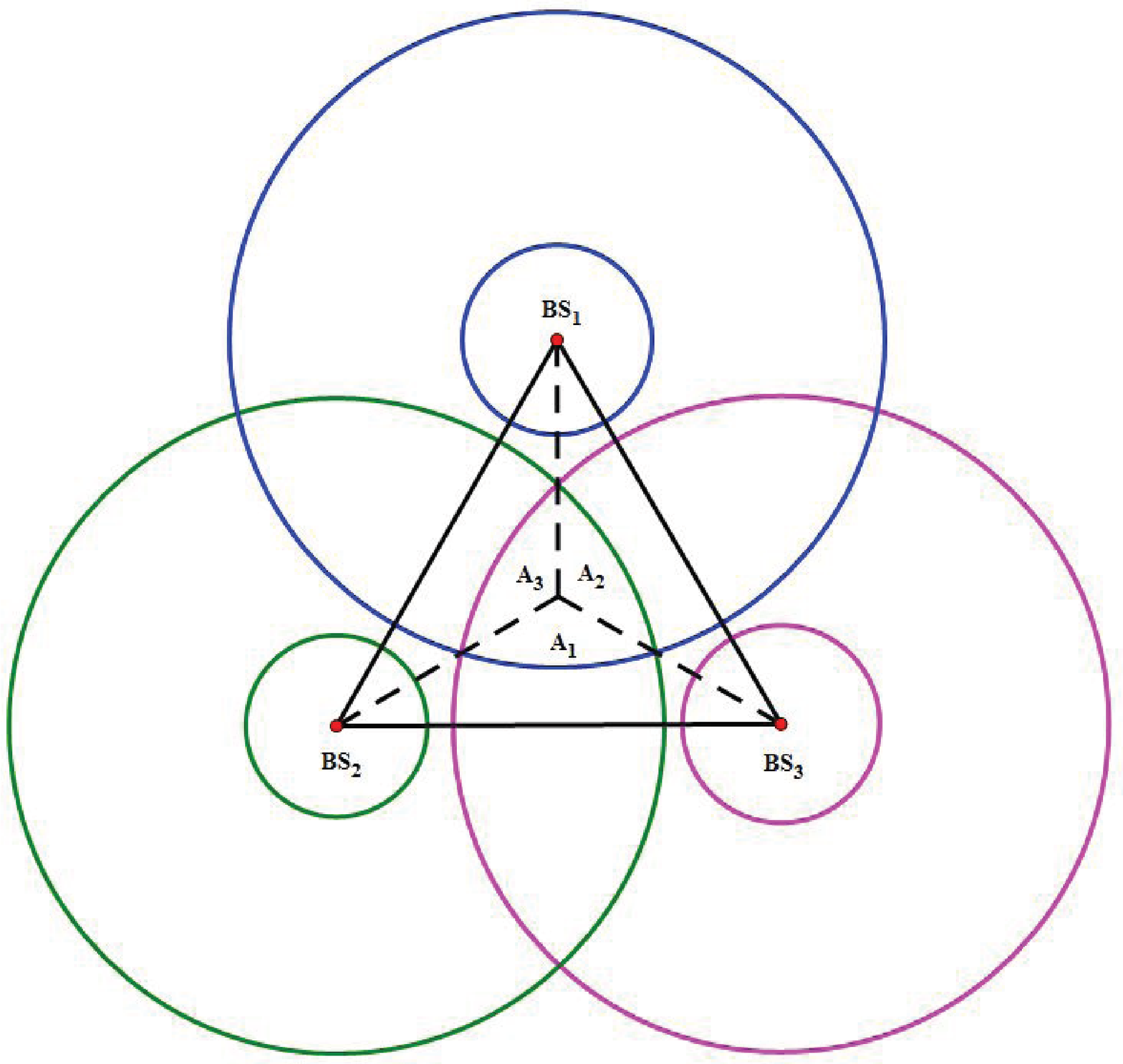}%
\label{system_b}}
\caption{System model}
\label{system}
\end{figure*}

Consider {\color{black}an} N-NOMA downlink communication scenario, in which three BSs are supporting a cell-edge user cooperatively, while each BS is also individually communicating with a {\color{black}close} user. All the nodes in the considered scenario {\color{black}are equipped with} a single antenna. More specifically, as shown in Fig. \ref{system}, consider an equilateral triangle, whose side length is denoted by $l$. Each BS, denoted by BS i, $1\leq i\leq3$, is located at one of the vertex of the equilateral triangle. There is a big disc and a small disc centered at BS i, respectively. The big one is denoted by $D_{0i}$, and the small one by $D_i$. The locations of the users are described as follows. The cell-edge user is denoted by user 0 and its position (denoted by $p_0$) is assumed to be uniformly distributed in the the intersecting area ({\color{black}which is} denoted by $A$, {\color{black}and} composed of three symmetrical areas, $A_1,A_2$ and $A_3$) of the three big discs. The user close to BS i is denoted by user i and its position (denoted by $p_i$) is assumed to be uniformly distributed in disc $D_i$. In addition, the size of the discs {\color{black}can be} described as follows: the three big discs are assumed to have the same radius which is denoted by $R_0$; the radius of each small disc, i.e., $D_i$, $1\leq i\leq3$, is denoted by $R_i$.

{\color{blue} A practical assumption about the range of $R_0$ is $\frac{\sqrt{3}}{3}l \leq R_0\leq\frac{\sqrt{3}}{2}l$. The reasons for making this assumption are listed as follow. On the one hand, if $R_0$ is too small, the three big discs with radius $R_0$ have no intersecting area. On the other hand, if $R_0$ is too big, user 0 may be located too close to a BS, which contradicts with the assumption that user 0 is a cell edge user. Besides the above assumption for the big disc, the radius of each small disc should be carefully chosen to ensure that the cell-edge user is much further from the BS than the near users to apply NOMA.}

The channel {\color{black}modeling used} in this paper {\color{black}is described} as follows. The channel gain between BS i and user j is modeled as  $h_{ij}=g_{ij}/\sqrt{L_{ij}}$, where $g_{ij}$ is the Rayleigh fading gain, i.e., $g_{ij}\sim CN\left(0,1\right)$, $L_{ij}=d_{ij}^\alpha$ denotes the path loss, $d_{ij}$ is the distance between BS i and user j, {\color{blue}and $\alpha$ is the path loss exponent parameter.}

It is important to point out that, in the considered N-NOMA scheme, at the transmitter, BS i only needs to know the phase information of the channel between BS i and user 0, while at the receiver side, {\color{black}the} users have the full CSI.

BS i sends the following information
\begin{align}\label{transmit}
x_i = \frac{h_{i0}^{*}}{|h_{i0}|}\beta_0 \sqrt{P_s}s_0+\frac{h_{i0}^{*}}{|h_{i0}|}\beta_1 \sqrt{P_s}s_i,
\end{align}
where $s_i$, $0 \leq i \leq 3$, is the signal intended for user i, $E[|s_i|^2]=1$, $P_s$ is the transmit power, {\color{blue}and $\beta_0$, $\beta_1$ are the power allocation coefficients} {\color{black}with $ \beta_0^2+\beta_1^2=1$}. In this paper, $\beta_0$ and $\beta_1$ are set to be constant and the same {\color{black}in} all BSs. More sophisticated power allocation strategies can further improve the performance of the proposed N-NOMA system, but this is beyond the scope of this paper.

At the receiver side, the signal observed by user j,  $0\leq j\leq3$,  is given by
\begin{align}
  y_j &= \sum\limits_{i=1}^{3}{\left( \frac{h_{i0}^{*}h_{ij}}{|h_{i0}|}\beta_0 \sqrt{P_s} s_0+\frac{h_{i0}^{*}h_{ij}}{|h_{i0}|}\beta_1 \sqrt{P_s}s_i\right)}+n_j,
\end{align}
where $n_j$ is the noise observed at user j, and is modeled as a circular symmetric complex Gaussian random variable, i.e., $n_j\sim CN\left(0,\sigma^2\right)$, {\color{black}with $\sigma^2$ being} the noise power.

{\color{black}One can note that the expression for the near user's received signal is different from that for the cell-edge's received signal{\color{black}. M}ore specifically:}

{\color{black}a)} at user j, $1\leq j\leq3$, i.e., the near user, an interesting observation is that $\frac{h_{i0}^*}{|h_{i0}|}$ is uniformly distributed in $[0,2\pi]$ and it is independent from $h_{ij}$. Thus $\frac{h_{i0}^*}{|h_{i0}|}h_{ij}\overset{\Delta}{=}\tilde{h}_{ij}$ and $h_{ij}$ are identically distributed, and then the received signal can be rewritten as
\begin{align}
  y_j &= \sum\limits_{i=1}^{3}{ \left( \tilde{h}_{ij}\beta_0 \sqrt{P_s} s_0 + \tilde{h}_{ij}\beta_1 \sqrt{P_s}s_i \right)}+n_j,j=1,2,3.
\end{align}
According to the NOMA principle, user j, $1\leq j\leq3$, carries out successive interference cancellation (SIC) by first removing the message to user 0 with
\begin{align}
   \text{SINR}_{j,0}&=\frac{ \left| \sum\limits_{i=1}^{3}{\tilde{h}_{ij}}\right|^2\beta_0^2}
                {\sum\limits_{i=1}^{3}{\left| \tilde{h}_{ij}\right|^2}\beta_1^2+1/\rho},
\end{align}
where $\rho=P_s/\sigma^2$ is the transmit signal-to-noise ratio (SNR).
If successful, user j then decodes its own message with
\begin{align}
  \text{SINR}_{j} = \frac{ \left|\tilde{h}_{jj}\right|^2\beta_1^2}{\sum\limits_{\substack{i=1\\i \neq j}}^{3}{|\tilde{h}_{ij}|^2}\beta_1^2+1/\rho}.
\end{align}

{\color{black}b)} at user 0, i.e.,the cell-edge user, the received signal can be expressed as
\begin{align}
y_0 = \sum\limits_{i=1}^{3}{\left( |h_{i0}|\beta_0 \sqrt{P_s} s_0+|h_{i0}|\beta_1 \sqrt{P_s}s_i\right)}+n_0.
\end{align}
 {\color{black}In order to} decode the received message, user 0 treats $s_i$, $1\leq i\leq3$, as additive noise, which means that user 0 decodes its message with
\begin{align}
  \text{SINR}_0 &=\frac{\left( \sum\limits_{i=1}^{3}|h_{i0}| \right)^2\beta_0^2} {\sum\limits_{i=1}^{3}{|h_{i0}|^2}\beta_1^2+1/\rho}.
\end{align}

For comparison purposes, the following benchmark scheme, based on conventional OMA, is considered.  {\color{black}This} scheme only serves the cell-edge user by using the same channel resource block as the above N-NOMA scheme. Distributed digital beamforming is employed, i.e., the transmit signal at each BS is given by
\begin{align}
\tilde{x_i}=\frac{h_{i0}^*}{\sqrt{\sum\limits_{i=1}^{3}{|h_{i0}|^2}}}\sqrt{3P_s}s_0.
\end{align}
Note that, the overall transmit power of the three BSs is $3P_s$, which is the same as the proposed N-NOMA scheme. It is also important to point out that, in the benchmark OMA scheme, all the nodes have access to the full CSI.
\section{PERFORMANCE ANALYSIS}
In this section, we use the outage probability as the criterion to characterize the performance of the proposed N-NOMA scheme. Meanwhile, the outage probability achieved by the benchmark scheme is also obtained by considering the impact of the random location of the cell-edge user. {\color{blue}It should be pointed out that throughout the paper, the signals intended for different users are independently coded with Gaussian codebooks to achieve the Shannon capacity.}
\subsection{Outage performance at user 0}
The outage probability for user 0 to decode its information is given by
\begin{align}\label{P_0_pri}
 P_0&=P\left( {\frac{\left( \sum\limits_{i=1}^{3}|h_{i0}| \right)^2\beta_0^2}
                       {\sum\limits_{i=1}^{3}{|h_{i0}|^2}\beta_1^2+1/\rho}} < \eta_0 \right),
\end{align}
where $\eta_j=2^{r_j}-1$ and $r_j$ is the target data rate of user j, $0 \leq j \leq3$. It is worth rewriting $P_0$  {\color{black}as}
\begin{align}\label{P01}
 P_0&=P\left( \left( \beta_0^2-\beta_1^2\eta_0\right)\left(|h_{10}|^2+|h_{20}|^2+|h_{30}|^2\right)
                 +2\beta_0^2\left( |h_{10}||h_{20}|+|h_{10}||h_{30}|+|h_{20}||h_{30}|\right)
                 <\frac{\eta_0}{\rho}\right).
\end{align}
As {\color{black}it} can be seen in (\ref{P01}), the outage probability achieved by user 0 can be less than 1, even when  $\beta_0^2-\beta_1^2\eta_0\leq0$. But, in this paper, we assume that $\beta_0^2-\beta_1^2\eta_0>0$. The reason for making this assumption is that, when $\beta_0^2-\beta_1^2\eta_0\leq0$, the outage probabilities achieved by user 1,2  {\color{black}and} 3 will always be 1. Under this assumption, {\color{blue}it is not hard to find the impact of $l$ and $\beta_1$ on the outage performance of user 0 as shown in the following propositions.}
\begin{Proposition}
The outage probability achieved by user 0 is a monotonically increasing function with respect to $l$ (i.e.,the distance between the BSs), when $\beta_0, \eta_0, \rho$ and $k$ are fixed.
\end{Proposition}
\begin{Proposition}
The outage probability achieved by user 0 is a monotonically increasing function of $\beta_1$, which is the power allocation coefficient corresponding to the near user.
\end{Proposition}

The following lemma, which provides a closed-form expression for $P_0$, is  {\color{black}presented}.

\begin{lemma}
In the high SNR regime, the outage probability at user 0 can be {\color{blue}approximated} in closed-form as
\begin{align}\label{P_0}
P_0&\approx\underset{\kappa(\beta_0,\beta_1,\eta_0,\rho)}{\underbrace{\frac{4\left(G(\phi)-F(\phi)+F(0)\right)}{(\beta_0^2-\beta_1^2\eta_0)^2}}}E_{p_0}\left\{L_{10}L_{20}L_{30}\right\},
\end{align}
where $\phi=\sqrt{\frac{\eta_0}{\rho(2\beta_0^2-\beta_1^2\eta_0)}}$, the functions F(v) and G(v) are defined as
\begin{align}
\begin{cases}
F(v)=&\frac{2\sqrt{2}\beta_0^2}{48a^{3/2}}\left( -\frac{c^3 (2 a-b) \tanh ^{-1}\left(\frac{\sqrt{a} v}{\sqrt{(a+b)v^2+c}}\right)}{b^2}-\frac{\sqrt{a} v \sqrt{(a+b)v^2+c} \left(b
    (b-2 a)v^4+2 b c v^2+c^2\right)}{b}\right.\notag\\
   &+v^2 \left(b(4a+b)v^4 +3(2 a+b)cv^2 +3 c^2\right) \log
   \left(\frac{\sqrt{v^2 (a+b)+c}+\sqrt{a} v}{\sqrt{b v^2+c}}\right)\notag\\
   &\left.+\frac{2 a^{3/2} c^3 \log \left(\sqrt{a+b} \sqrt{ (a+b)v^2+c}+a v+b
   v\right)}{b^2 \sqrt{a+b}}\right)\\
G(v)=&\frac{2}{15} \left(\frac{a v^6}{6}+\frac{5 c v^4}{4}+\frac{5 d v^6}{6}\right),
\end{cases}
\end{align}
$a,b,c$ and $d$ are  {\color{black}used} to simplify  {\color{black}the expression} and are given by
\begin{align}
\begin{cases}
a=\beta_0^2\beta_1^2\eta_0-\beta_1^4\eta_0^2\\
b=3\beta_0^2\beta_1^2\eta_0-\beta_1^4\eta_0^2\\
c=(\beta_0^2-\beta_1^2\eta_0)\eta_0/\rho\\
d=2\beta_0^4+3\beta_0^2\beta_1^2\eta_0-\beta_1^4\eta_0^2\\
\phi=\sqrt{\frac{\eta_0}{\rho(2\beta_0^2-\beta_1^2\eta_0)}}\notag\\
\end{cases},
\end{align}
and $E_{p_0}\left\{L_{10}L_{20}L_{30}\right\}$, is the expectation of the product of $L_{10}$, $L_{20}$ and $L_{30}$ with respect to $p_0$,  when $\alpha=2$, $E_{p_0}\left\{L_{10}L_{20}L_{30}\right\}$ is given by
\begin{align}\label{E_L}
E_{p_0}\left\{L_{10}L_{20}L_{30}\right\}&=
\frac{1}{8S_{A}}\left(\frac{l^8}{8 \sqrt{3}}+\left(3 \sqrt{3}+4 \pi \right) l^4R_0^4+8 \left(\sqrt{3}+\pi \right) l^2 R_0^6+2 \pi R_0^8\right. \notag\\
&-6 R_0^4 \left(2 l^4+4 l^2 R_0^2+R_0^4\right) \sin
   ^{-1}\left(\frac{l}{2 R_0}\right)\notag\\
&\left.-\frac{1}{8} l R_0
   \sqrt{4-\frac{l^2}{R_0^2}} \left(l^6+2 l^4 R_0^2+102 l^2
   R_0^4+84 R_0^6\right)\right),
\end{align}
$S_{A}$ is the area of $A$, and can be expressed as
\begin{align}
   S_{A}=3R_0^2\left(\frac{\pi}{3}-\arcsin\left(\frac{l}{2R_0}\right)\right)-\sqrt{3}lR_0\sin{\left(\frac{\pi}{3}-\arcsin\left(\frac{l}{2R_0}\right)\right)},
\end{align}
{\color{blue}when $\alpha>2$, $E_{p_0}\left\{L_{10}L_{20}L_{30}\right\}$ is approximated by considering the special case when user $0$ is located very close to the center of the equilateral triangle and given by
\begin{align}\label{E_L2}
E_{p_0}\left\{L_{10}L_{20}L_{30}\right\}&\approx 3^{-\frac{3\alpha}{2}}l^{3\alpha}.
\end{align}}
\end{lemma}
\begin{IEEEproof}
Please refer to Appendix A.
\end{IEEEproof}
{\color{blue}\begin{Remark}
The main difficulties to get $P_0$ are listed as follows: firstly, the distributions of $|h_{i0}|$ are correlated because $|h_{i0}|$ is revelent to the location of user $0$. Secondly, in (\ref{P01}), the left hand side of the inequality contains product terms. As can be seen in Appendix A, by considering the high SNR regime, the mentioned two difficulties be can be solved separately for analysis simplification.
\end{Remark}}

{\color{black}As shown in Lemma 1, the first part of the expression for $P_0$, $\kappa(\beta_0, \beta_1, \eta_0, \rho)$,  is a function {\color{black}of} $\beta_0,\beta_1,\eta_0$ and $\rho$, while the second part,  $E_{p_0}\left\{L_{10}L_{20}L_{30}\right\}$, is a function {\color{black}of} $l$ and $R_0$. By further analyzing $\kappa(\beta_0,\beta_1,\eta_0,\rho)$, it is not hard to {\color{black}conclude} that the diversity obtained by user 0 is 3, since there is a common factor $1/\rho^3$.

To get insight into the impact of $R_0$ on $P_0$, we focus on the case when $\alpha=2$. Note that, (\ref{E_L}) can be rewritten as
\begin{align}
E_{p_0}\left\{L_{10}L_{20}L_{30}\right\}=\lambda(k)l^6,
\end{align}
where $k=\frac{R_0}{l}$, and $\lambda(k)$ is a function of $k$, which is expressed as
\begin{align}\label{lambda_k}
\lambda(k)=&\frac{1}{192k \left(\pi  k-3 k \sin ^{-1}\left(\frac{1}{2k}\right)-\sqrt{3} \cos \left(\sin^{-1}\left(\frac{1}{2 k}\right)+\frac{\pi}{6}\right)\right)}\times\notag\\
&\left(48 \pi  k^8-3 \sqrt{4-\frac{1}{k^2}} \left(84
   k^6+102 k^4+2 k^2+1\right) k+\sqrt{3}+\right.\notag\\
&\left.192 \left(\sqrt{3}+\pi \right) k^6+24 \left(3 \sqrt{3}+4 \pi
   \right) k^4-144 \left(k^4+4 k^2+2\right) k^4 \sin
   ^{-1}\left(\frac{1}{2 k}\right)\right).
\end{align}

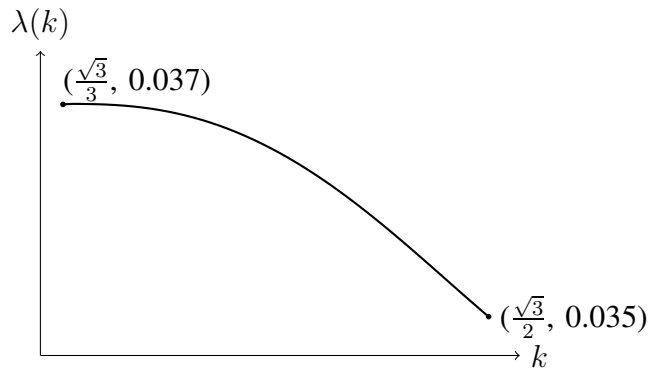
\begin{figure}[h]
\vspace{-1em}
\setlength{\belowcaptionskip}{-1em}   
 \centering
  \begin{tikzpicture}
    \draw[->] (3^0.5/3*20-0.3,0.0348*1500) -- (3^0.5/2*20+0.3,0.0348*1500) node[right] {$k$};
    \draw[->] (3^0.5/3*20-0.3,0.0348*1500) -- (3^0.5/3*20-0.3,0.0375*1500) node[above] {$\lambda(k)$};
    \draw[thick] (11.547017,55.548502) --
(11.662487,55.555470) --
(11.777957,55.554903) --
(11.893427,55.553438) --
(12.008897,55.550724) --
(12.124367,55.546457) --
(12.239837,55.540376) --
(12.355307,55.532252) --
(12.470777,55.521888) --
(12.586247,55.509110) --
(12.701717,55.493766) --
(12.817188,55.475723) --
(12.932658,55.454865) --
(13.048128,55.431092) --
(13.163598,55.404317) --
(13.279068,55.374466) --
(13.394538,55.341477) --
(13.510008,55.305301) --
(13.625478,55.265898) --
(13.740948,55.223238) --
(13.856418,55.177305) --
(13.971888,55.128088) --
(14.087358,55.075590) --
(14.202828,55.019821) --
(14.318298,54.960805) --
(14.433768,54.898572) --
(14.549238,54.833163) --
(14.664708,54.764632) --
(14.780178,54.693042) --
(14.895648,54.618465) --
(15.011119,54.540987) --
(15.126589,54.460705) --
(15.242059,54.377726) --
(15.357529,54.292171) --
(15.472999,54.204173) --
(15.588469,54.113877) --
(15.703939,54.021443) --
(15.819409,53.927043) --
(15.934879,53.830865) --
(16.050349,53.733111) --
(16.165819,53.633997) --
(16.281289,53.533757) --
(16.396759,53.432639) --
(16.512229,53.330908) --
(16.627699,53.228849) --
(16.743169,53.126762) --
(16.858639,53.024965) --
(16.974109,52.923798) --
(17.089580,52.823619) --
(17.205050,52.724804) ;
\draw[fill] (11.547017,55.548502) circle [radius=0.03];
\draw[fill] (17.205050,52.724804) circle [radius=0.03];
\node [above,right] at (11.4,55.88502) {($\frac{\sqrt{3}}{3}$, 0.037)};
\node [right] at (17.205050,52.724804) {($\frac{\sqrt{3}}{2}$, 0.035)};
\end{tikzpicture}
\caption{Illustration of function $\lambda(k)$}
\label{lambda}
\end{figure}

The simulation results shown in  Fig. \ref{lambda}  indicate that $\lambda(k)$ is a monotonically decreasing function of k in the interval of $\frac{\sqrt{3}}{3} \leq k \leq \frac{\sqrt{3}}{2}$, although we cannot find a formal proof for this yet. Thus we {\color{black}highlight} the following remark.

\begin{Remark}
In the high SNR regime, the outage probability achieved by user 0 is a monotonically decreasing function with respect to k (or $R_0$), when $\beta_0,\eta_0, \rho$ and $l$ are fixed.
\end{Remark}

It is also interesting to consider the special case when user 0 is located very close to $O$. Fig. \ref{lambda} indicates that, when k approaches $\frac{\sqrt{3}}{3}$ from the right-hand side, the slope of the curve is nearly 0. Thus when user 0 is located very close to $O$, the impact of {\color{black}different choices of $R_0$}  on $P_0$ is negligible, this observation is consistent with the expression (\ref{E_L2}) presented in Lemma 1, where $R_0$ is neglected.

\subsubsection{The case of the benchmark network OMA scheme}
To facilitate comparison, we also consider the special case when user 0 is located very close to $O$ for the benchmark OMA scheme. The following lemma characterize the outage performance of the benchmark OMA scheme:
{\color{blue}\begin{lemma}
In the scenario where user 0 is located very close to $O$, the outage probability achieved by user 0 over the benchmark OMA scheme can be approxiamted as follows:
\begin{align}\label{P_OMA}
\tilde{P_0}&\approx\frac{ e^{-\frac{3^{-\frac{\alpha }{2}-1} \eta_0
    l^{\alpha }}{\rho }} \left(-\eta_0 ^2 l^{2 \alpha }+2\times
   3^{\alpha +2} \rho ^2 \left(e^{\frac{3^{-\frac{\alpha
   }{2}-1} \eta_0  l^{\alpha }}{\rho }}-1\right)-2\times
   3^{\frac{\alpha }{2}+1} \eta_0  \rho  l^{\alpha }\right)}{2\times3^{\alpha+2}
   \rho ^2}
\end{align}
\end{lemma}}
\begin{IEEEproof}
Please refer to Appendix B.
\end{IEEEproof}
\subsection{Outage performance at user j, $1\leq j\leq3 $, achieved by N-NOMA }
The QoS requirement of user j can be  {\color{black}met} only when the following two constraints are satisfied:  {\color{black}a)} user j can decode the message intended to user 0,  and  {\color{black}b)} user j can decode its own message after successfully removing the message intended to user 0. More rigorously, we define the outage events at user j as follows. First define,  $E_{j,k}\overset{\Delta}{=}\left\{\text{SINR}_{j,k}<\eta_k\right\}$, as the event that user j cannot decode the message intended to user k, where $1\leq j\leq3$, $k\in \left\{0,j\right\}$, and
$E_{j,k}^{c}$ as the complementary set of $E_{j,k}$. Thus the outage probability at user j can be expressed as
\begin{align}
P_j&=1-P(E_{j,0}^c\cap E_{j,j}^c)=P(E_{j,0}\cup E_{j,j}).
\end{align}
{\color{black}In order to} reduce the complexity of the performance analysis, it is important to note that,  when $l>>R_j$, $\tilde{h}_{ij}$ $(i\neq j)$ is much smaller than $\tilde{h}_{jj}$ because of large scale propagation losses.  Thus the SINR to decode $s_0$ at user j, $1\leq j\leq3$, can be approximated as
\begin{align}
\text{SINR}_{j0}&\approx\frac{ \left| \tilde{h}_{jj}\right|^2\beta_0^2}
                {\sum\limits_{i=1}^{3}{\left| \tilde{h}_{ij}\right|^2}\beta_1^2+1/\rho}.
\end{align}
{\color{blue}This assumption is reasonable, since in realistic systems, the distances between the BSs are usually much further than that between a BS and its near user.}
Following this approximation, $E_{j,0}$ can be expressed as
\begin{align}\label{E_j0}
E_{j,0}=\left\{\left| \tilde{h}_{jj}\right|^2 < \frac{\eta_0}{\beta_0^2-\beta_1^2\eta_0}\left(\sum\limits_{\substack{i=1\\i \neq j}}^{3}{\left| \tilde{h}_{ij}\right|^2}\beta_1^2+1/\rho\right)\right\}.
\end{align}
It is important to point out that the assumption $\beta_0^2-\beta_1^2\eta_0>0$ is applied in the above expression. Since if $\beta_0^2-\beta_1^2\eta_0 \leq 0$, $P_j$ will always be 1. Using the same format as in (\ref{E_j0}), $E_{j,j}$ can be expressed as
\begin{align}\label{E_jj}
E_{j,j}=\left\{\left| \tilde{h}_{jj}\right|^2 < \frac{\eta_j}{\beta_1^2}\left(\sum\limits_{\substack{i=1\\i \neq j}}^{3}{\left| \tilde{h}_{ij}\right|^2}\beta_1^2+1/\rho\right)\right\}.
\end{align}
Therefore the outage probability achieved by user j can be further  {\color{black}formulated} as
\begin{align}\label{P_j_23}
P_j=P\left(\left| \tilde{h}_{jj}\right|^2 < M_j\left(\sum\limits_{\substack{i=1\\i \neq j}}^{3}{\left| \tilde{h}_{ij}\right|^2}\beta_1^2+1/\rho\right) \right),
\end{align}
where $M_j=\max\left\{\frac{\eta_0}{\beta_0^2-\beta_1^2\eta_0},\frac{\eta_j}{\beta_1^2}\right\}$. The following lemma characterizes the outage performance at user j, $1 \leq j\leq3$.
\begin{lemma}
When $l>>R_j$, {\color{black}then} the outage probability achieved by user j, $1\leq j\leq3$, can be {\color{blue}approximated} as{\color{blue}
\begin{align}\label{P_j}
P_j\approx1-\frac{2\rho^\frac{2}{\alpha}M_j\gamma(\frac{2}{\alpha},\frac{M_jR_j^\alpha}{\rho})
  -\frac{4\beta_1^2M_j\rho^{\frac{2}{\alpha}+1}}{l^\alpha}\gamma(\frac{2}{\alpha}+1,\frac{M_jR_j^\alpha}{\rho})}{\alpha M_j^{\frac{2}{\alpha}+1}R_j^2}.
\end{align}
where $\gamma(s,x)$ is the the lower incomplete gamma function defined as $\gamma(s,x) = \int_0^x t^{s-1}\,e^{-t}\,{\rm d}t$.}
\end{lemma}
\begin{IEEEproof}
Please refer to Appendix C.
\end{IEEEproof}

By using  Lemma 3, the following corollaries can be obtained to show how system parameters, $\beta_1$, $l$ and $R_j$, affect the outage performance, achieved by user j.

\begin{corollary}
When $l>>R_j$, {\color{black}then} the outage probability {\color{black}at high SNR} achieved by user j, $1\leq j\leq3$, is a monotonically increasing function of $R_j$.
\end{corollary}
\begin{IEEEproof}
{\color{blue}Here, we concentrate on the case when $\alpha=2$. Note that, from (\ref{P_j}), for $\alpha=2$, $P_j$ can be approximated as
\begin{align}
P_j\approx1-\frac{\rho  e^{-\frac{M_j R_j^2}{\rho }}
   \left(l^2 \left(e^{\frac{M_j R_j^2}{\rho
   }}-1\right)+2 \beta_1 ^2 \left(-\rho
   e^{\frac{M_j R_j^2}{\rho }}+M_j R_j^2+\rho
   \right)\right)}{l^2 M_j R_j^2}.
\end{align}}
Note that, the partial derivative of $R_j^2$ for $P_j$ is
\begin{align}
\frac{\partial{P_j}}{\partial{R_j^2}}=\frac{M_j \left(3 l^2+2 \beta_1 ^2 \left(3\rho -4 M_j R_j^2\right)\right)}{6 l^2 \rho}.
\end{align}
Note that $l>>R_j$, thus $\frac{\partial{P_j}}{\partial{R_j^2}}>0$ and {\color{black}the proof is completed}.
\end{IEEEproof}
\begin{corollary}
When $l>>R_j$, {\color{black}then}
\begin{enumerate}
  \item if $0<\beta_1^2 \leq \frac{\eta_j}{\eta_0+\eta_j+\eta_0\eta_j}$, the outage probability achieved by user j, $1\leq j\leq3$, is a monotonically decreasing function of $\beta_1^2$,
  \item if $\frac{\eta_j}{\eta_0+\eta_j+\eta_0\eta_j}<\beta_1^2<\frac{1}{1+\eta_0}$, the outage probability achieved by user j, $1\leq j\leq3$, is a monotonically increasing function of $\beta_1^2$.
\end{enumerate}
\end{corollary}
\begin{IEEEproof}
\begin{enumerate}
\item
When $0<\beta_1^2\leq\frac{\eta_j}{\eta_0+\eta_j+\eta_0\eta_j}$ and $M_j=\frac{\eta_j}{\beta_1^2}$, $P_j$ can be expressed as
\begin{align}
P_j=P\left(\left| \tilde{h}_{jj}\right|^2 < \frac{\eta_j}{\beta_1^2}\left(\sum\limits_{\substack{i=1\\i \neq j}}^{3}{\left| \tilde{h}_{ij}\right|^2}\beta_1^2+1/\rho\right) \right).
\end{align}
Note that $\frac{\eta_j}{\beta_1^2}\left(\sum\limits_{\substack{i=1\\i \neq j}}^{3}{\left| \tilde{h}_{ij}\right|^2}\beta_1^2+1/\rho\right)$ in (\ref{P_j_23}) is a monotonically decreasing function of $\beta_1^2$, thus $P_j$ is {\color{black}also} a monotonically decreasing function of $\beta_1^2$.
\item
When $\frac{\eta_j}{\eta_0+\eta_j+\eta_0\eta_j}<\beta_1^2<\frac{1}{1+\eta_0}$ and  $M_j=\frac{\eta_0}{\beta_0^2-\beta_1^2\eta_0}$, $P_j$ can be expressed as
\begin{align}
P_j=P\left(\left| \tilde{h}_{jj}\right|^2 < \frac{\eta_0}{\beta_0^2-\beta_1^2\eta_0}\left(\sum\limits_{\substack{i=1\\i \neq j}}^{3}{\left| \tilde{h}_{ij}\right|^2}\beta_1^2+1/\rho\right) \right).
\end{align}
Note that $\frac{\eta_0}{\beta_0^2-\beta_1^2\eta_0}\left(\sum\limits_{\substack{i=1\\i \neq j}}^{3}{\left| \tilde{h}_{ij}\right|^2}\beta_1^2+1/\rho\right)$ in (\ref{P_j_23}) is a monotonically increasing function of $\beta_1^2$, thus $P_j$ is {\color{black}also} a monotonically increasing function of $\beta_1^2$.
\end{enumerate}
\end{IEEEproof}
{\color{blue}\subsection{Outage performance at user j, $1\leq j\leq3 $, achieved by N-NOMA with co-channel interference}
Until now, the outage performance of the cell-edge user and the near users are analyzed by ignoring the co-channel interference outside the considered three BSs. However, in a practical system, the considered users receive signals not only from the mentioned BSs, but also from co-channel interference sources, hence, it is essential to consider the case with co-channel interference. In this paper, we model the interference sources as a homogeneous Poisson point process $\Phi_I$ with intensity $\lambda_I$, i.e., $\Phi_I=\{{p_I}_k\}$, where ${p_I}_k$ is the location of the $k$-th interference source. For tractable analysis, it is assumed that the interference sources use identical transmission powers, denoted by $P_I$.

Under the above interference model, the signal observed by user j,  $0\leq j\leq3$, is now expressed as:
\begin{align}
  y_j &= \sum\limits_{i=1}^{3}{\left( \frac{h_{i0}^{*}h_{ij}}{|h_{i0}|}\beta_0 \sqrt{P_s} s_0+\frac{h_{i0}^{*}h_{ij}}{|h_{i0}|}\beta_1 \sqrt{P_s}s_i\right)}+\omega_{I_j}+n_j,
\end{align}
where $\omega_{I_j}$ is the overall co-channel interference observed by user j, and can be expressed as follows:
\begin{align}
\omega_{I_j}\overset{\Delta}{=}\sum\limits_{p_{I_k}\in\Phi_I}\frac{g_{I_k,j}}{\sqrt{L(d_{I_k,j})}}\sqrt{P_I}\tilde{s}_{I_k}
\end{align}
where $\tilde{s}_{I_k}$ is the normalized signal sent by the $k$-th interference source, i.e., $E[|\tilde{s}_{I_k}|^2]=1$, and these signals are assumed to be independent from each other. $d_{I_k,j}$ is the distance between user $j$ and the $k$-th interference source, and $g_{I_k,j}$ is the corresponding Rayleigh fading gain, i.e., $g_{I_k,j}\sim CN\left(0,1\right)$.

After considering the case with co-channel interference, the outage probabilities expressed in (\ref{P_0_pri}) and (\ref{P_j_23}) become
\begin{align}\label{P_0_PPP_pri}
 P_0^{\text{Inter}}&=P\left( {\frac{\left( \sum\limits_{i=1}^{3}|h_{i0}| \right)^2\beta_0^2}
                       {\sum\limits_{i=1}^{3}{|h_{i0}|^2}\beta_1^2+I_0+1/\rho}} < \eta_0 \right),
\end{align}
for the cell-edge user and
\begin{align}\label{P_j_PPP_pri}
P_j^{\text{Inter}}=P\left(\left| \tilde{h}_{jj}\right|^2 < M_j\left(\sum\limits_{\substack{i=1\\i \neq j}}^{3}{\left| \tilde{h}_{ij}\right|^2}\beta_1^2+I_j+1/\rho\right) \right), j=1,2,3
\end{align}
for near users, where $I_j=\sum\limits_{p_{I_k}\in\Phi_I}\frac{|g_{I_k,j}|^2}{L(d_{I_k,j})}\rho_I, 0\leq j\leq3$, where $\rho_I=\frac{P_I}{P_s}$. Analyzing the outage probability in (\ref{P_0_PPP_pri}) is very difficult, thus we focus on analyzing the outage probability for the near users. Thanks to stochastic geometry, we get the following lemma to characterize the outage performance of the near users in the case with co-channel interference.
\begin{lemma}
When $l>>R_j$, then the outage probability achieved by user j, $1\leq j\leq3$, can be {\color{blue}approxiamted} as
\begin{align}\label{P_j_inter}
P_j^{\text{Inter}}\approx\sum\limits_{n=1}^{N}\frac{\pi}{NR_j}\sqrt{1-\theta_n^2}f\left(\frac{R_j}{2}\theta_n+\frac{R_j}{2}\right),
\end{align}
where $N$ denotes the parameter for Gauss-Chebyshev quadrature, $\theta_n=\cos{\frac{(2n-1)\pi}{2N}}$ and
\begin{align}
f(x)=(x-\frac{2\beta_1^2M_j}{l^\alpha}x^{\alpha+1})
\exp\left(-\frac{M_j}{\rho}x^\alpha
-2\pi\lambda_I\frac{(M_j\rho_I)^{\frac{2}{\alpha}}}{\alpha}B\left(\frac{2}{\alpha},1-\frac{2}{\alpha}\right)x^2\right),
\end{align}
where $B(\cdot)$ is the Beta function.
\end{lemma}
\begin{IEEEproof}
Please refer to Appendix D.
\end{IEEEproof}
}
\section{Numerical Results {\color{black}and Simulations}}
\begin{figure*}[!t]
\vspace{-1em}
\centering
\subfloat[Outage Sum Rate]{\includegraphics[width=3.2in]{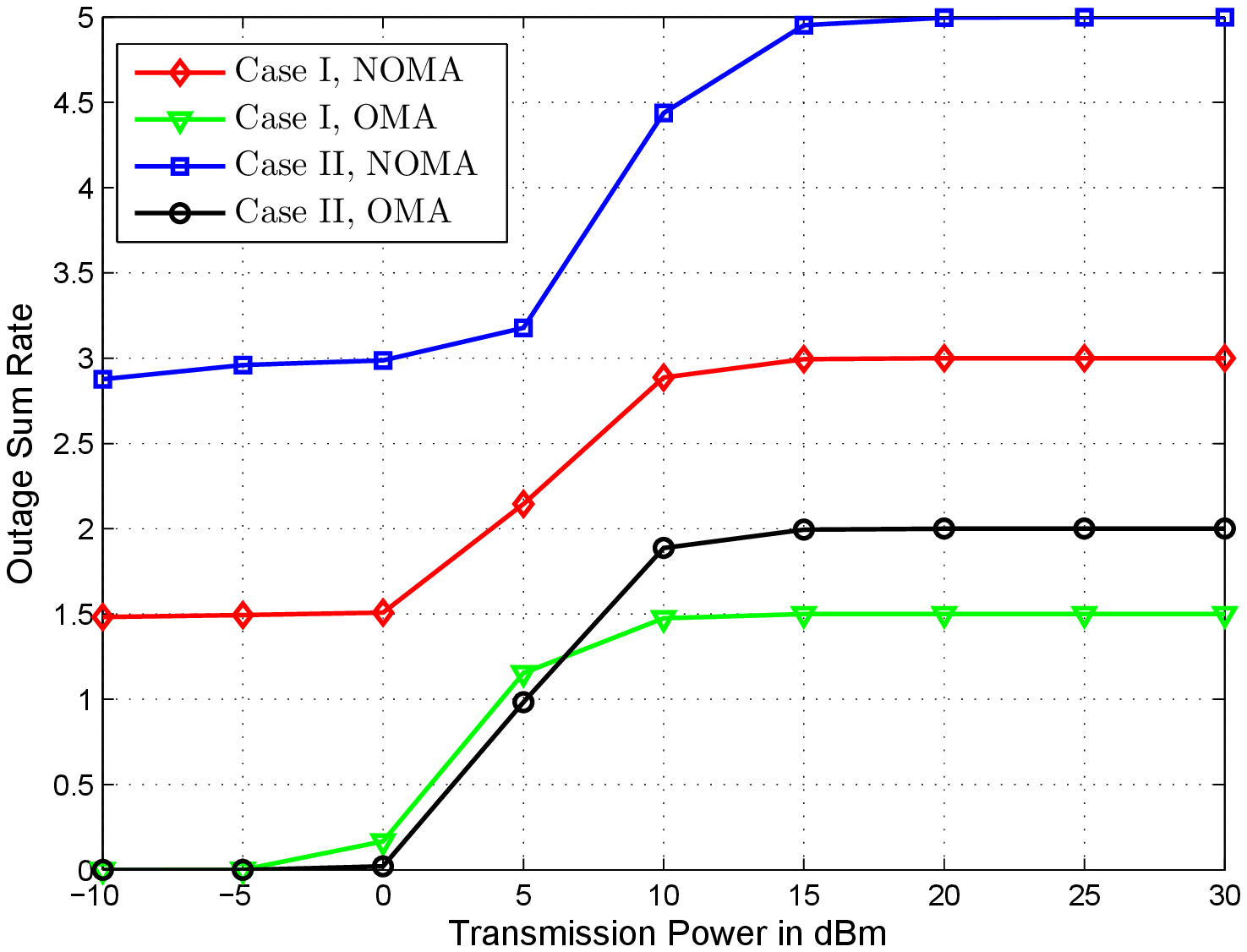}%
\label{compare_a}}
\hfil
\subfloat[Outage Probability]{\includegraphics[width=3.2in]{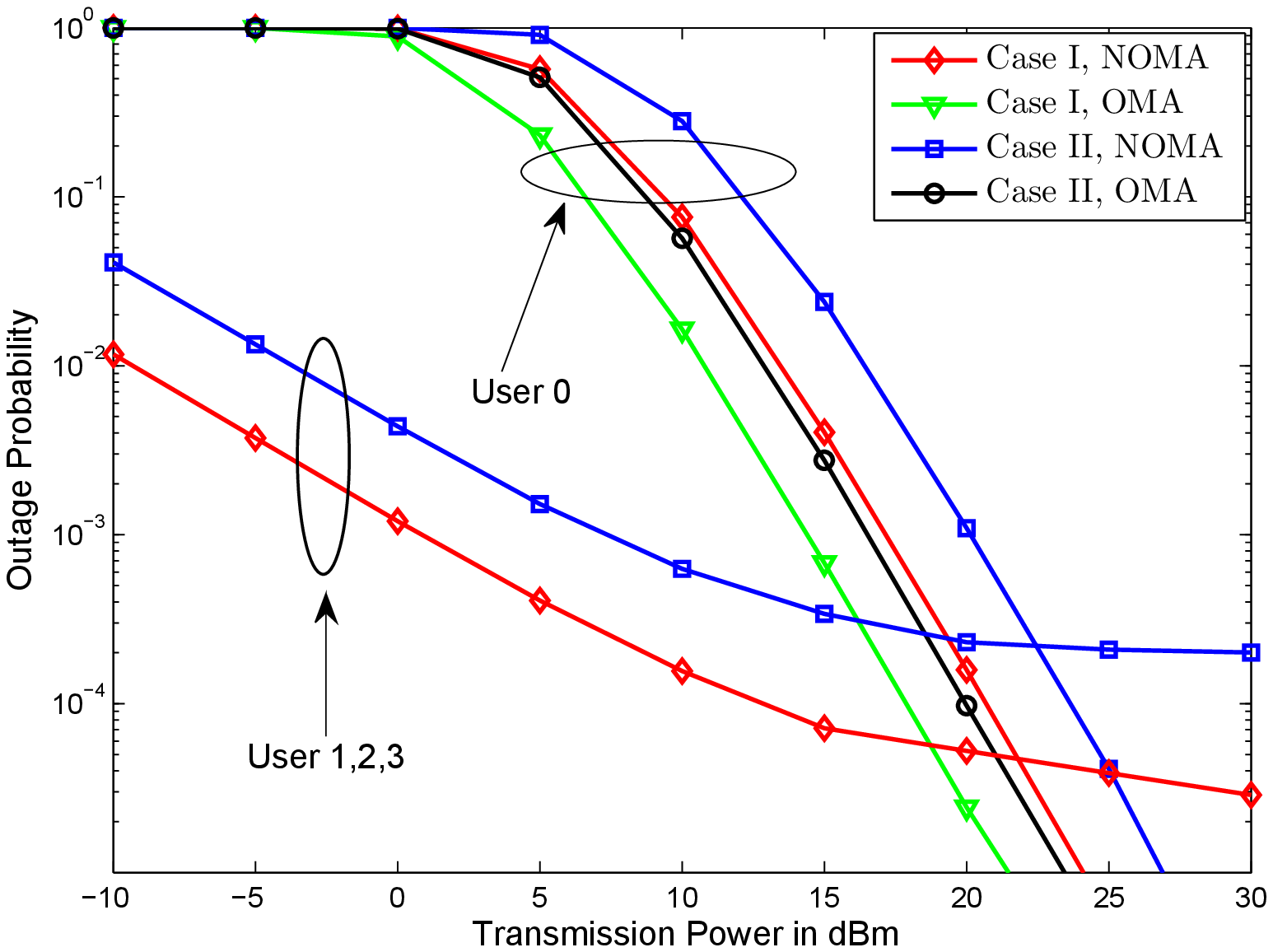}%
\label{compare_b}}
\caption{Performance comparison of N-NOMA and conventional OMA. Case I: $r_0=1.5$ BPCU, $r_1=r_2=r_3=0.5$ BPCU.  Case II: $r_0=2$ BPCU, $r_1=r_2=r_3=1$ BPCU.}
\label{compare}
\end{figure*}

\begin{figure}[!t]
\vspace{-1em}
\setlength{\belowcaptionskip}{-1em}   
\centering
\includegraphics[width=3.5in]{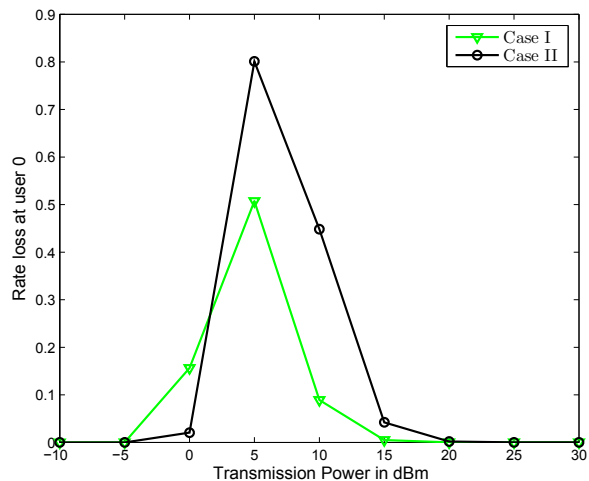}
\caption{Rate loss at the cell-edge user. Case I: $r_0=1.5$ BPCU, $r_1=r_2=r_3=0.5$ BPCU.  Case II: $r_0=2$ BPCU, $r_1=r_2=r_3=1$ BPCU.}
\label{compare_loss}
\end{figure}
In this section, computer simulations {\color{black}are performed} to demonstrate the performance of the proposed N-NOMA system and also verify the accuracy of the analytical results. {\color{blue}The thermal noise power is set as $-170$ dBm/Hz, the carrier frequency is $2\times10^9$ Hz, the transmission bandwidth is $10$ MHz, and the transmitter and receiver antenna gain are set as 1}.

Fig. \ref{compare} and Fig. \ref{compare_loss} shows a performance comparison of the proposed N-NOMA scheme and the conventional OMA scheme. In Fig. \ref{compare}(a), the outage sum rates are shown as functions of the transmission power, the corresponding outage probabilities are shown in Fig. \ref{compare}(b). {\color{blue}Note that, when using the N-NOMA, the cell-edge user may have some rate loss compared to the conventional OMA, this rate loss is quantified as shown in Fig. \ref{compare_loss}}.
{\color{black}The parameters are set as: $l=400m$, $R_0=250m$, $R_j=10m, 1\leq j\leq3$, $\beta_0^2=\frac{4}{5}, \alpha=3$.}
Fig. \ref{compare}(b) shows that, {\color{black}at user 0, the outage probability achieved by the conventional OMA is lower than that achieved by the considered N-NOMA.} The reason is that, the conventional OMA scheme serves only the cell-edge user and digital beamforming is superior to analog beamforming. But, as {\color{black}it} can be seen from Fig. \ref{compare}(a) and Fig. \ref{compare_loss}, with some tolerable rate loss at the cell edge user, the proposed N-NOMA scheme can achieve a {\color{black}higher} outage sum rate, compared to the conventional OMA,  which demonstrates the superior special efficiency of NOMA. For example, as shown in Fig. \ref{compare}(a), in Case II, when the transmit power is {\color{blue}30 dBm}, the sum rate {\color{black}achieved} by the proposed N-NOMA is about {\color{blue}5 bits} per channel use (BPCU), while {\color{black}that of} the conventional OMA scheme is about 2 BPCU. {\color{black}Hence} the gap is {\color{black}about} {\color{blue}3 BPCU}.
\begin{figure}[!t]
\vspace{-1em}
\setlength{\belowcaptionskip}{-1em}   
\centering
\includegraphics[width=3.5in]{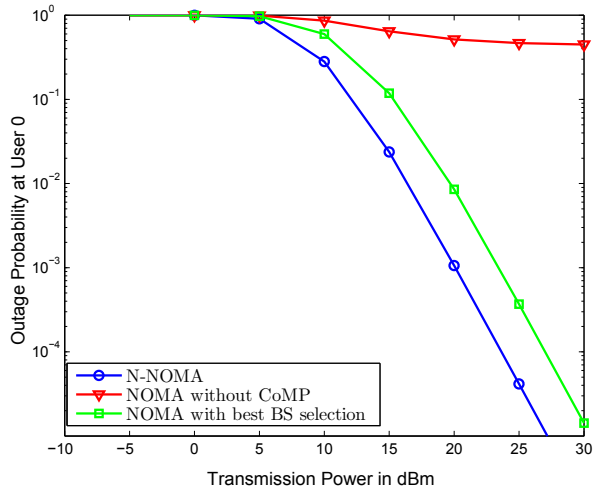}
\caption{Comparison of N-NOMA with NOMA without CoMP and NOMA with best BS selection.}
\label{compare_3NOMA}
\end{figure}

{\color{blue}Fig. \ref{compare_3NOMA}, shows a performance comparison of the N-NOMA with other two benchmark NOMA schemes named the NOMA without CoMP and the NOMA with best BS selection. In the two benchmark NOMA schemes, only one BS, say BS 1, employs NOMA to support the cell-edge user and its near user, that's to say, the transmit signals are as follows: $x_1 =\beta_0\sqrt{3P_s}s_0+\beta_1\sqrt{P_s}s_1$ and $x_j =\beta\sqrt{P_s}s_j, j=2,3$.
In the NOMA without CoMP scheme, the BS which serves the cell-edge user is randomly chosen from the three BSs, whereas in the NOMA with best BS selection scheme, the BS with largest channel gain, i.e., $|h_{i0}|^2$, is chosen. Note that, the transmission power for user $0$ becomes $3\beta_0^2P_s$ in the two benchmark NOMA schemes, this may be impractical for that the transmission power at BS 1 may excess the power constraint. Even so, as shown in Fig. \ref{compare_3NOMA}, the outage performance at user $0$ achieved by the N-NOMA is much better than that achieved by the two benchmark NOMA schemes.  The above observation indicates the importance of applying diversity techniques when serving the cell-edge user.
}
\begin{figure}[!t]
\vspace{-1em}
\setlength{\belowcaptionskip}{-1em}   
\centering
\includegraphics[width=3.5in]{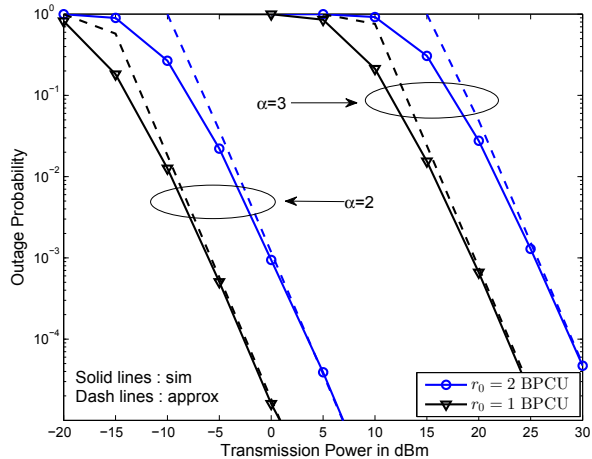}
\caption{Outage performance at user $0$ for the N-NOMA system. $l=600m$, $R_0=400m$.}
\label{accuarcy_0}
\end{figure}
\begin{figure}[!t]
\vspace{-1em}
\setlength{\belowcaptionskip}{-1em}   
\centering
\includegraphics[width=3.5in]{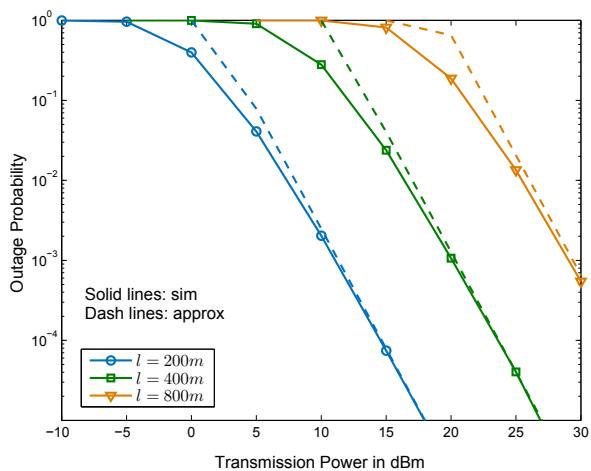}
\caption{Impact of $l$ on $P_0$. $\alpha=3$, $R_0=\frac{11}{10}\times\frac{l}{\sqrt{3}}$, $r_0=2$ BPCU.}
\label{impact_l}
\end{figure}

{\color{blue}In Fig. \ref{accuarcy_0} and Fig. \ref{impact_l}, the accuracy of the analytical results developed in Lemma 1 are verified.
{\color{blue}The power allocation coefficient is set as: $\beta_0^2=\frac{4}{5}$.}
As {\color{blue}it is evident} from Fig. \ref{accuarcy_0} and Fig. \ref{impact_l}, the approximation results {\color{blue}of} Lemma 1 are accurate in the high SNR regime. It is also verified in Fig. \ref{impact_l} that $P_0$ increases with $l$.}

\begin{figure}[!t]
\setlength{\belowcaptionskip}{-1em}   
\centering
\includegraphics[width=3.5in]{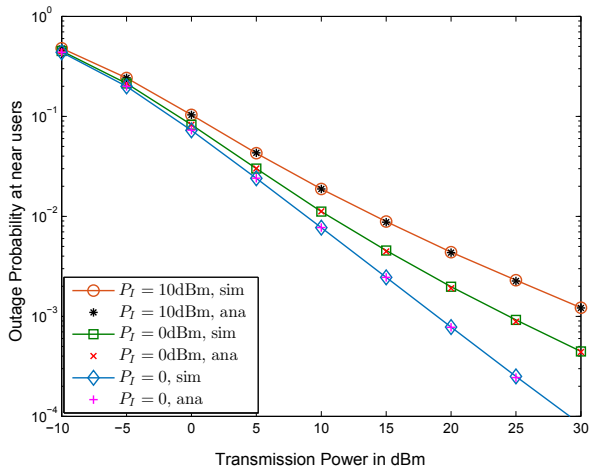}
\caption{Outage performance at near users.}
\label{user_123}
\end{figure}

{\color{blue}In Fig. \ref{user_123}, the accuracy of the analytical results developed in Lemma 3 and Lemma 4 are verified. The parameters are set as: $l=300m$, $R_0=200m$, $R_j=20m, 1\leq j\leq3$, $\beta_0^2=\frac{4}{5}$, $\lambda_I=\frac{1}{\pi200^2}$, $\alpha=4$, and $r_j=0.5$ BPCU, $0\leq j\leq3$. Note that, for the interference free case, i.e., $P_I=0$, the analytical results are based on Lemma 3, while for $P_I>0$, the analytical results are based on Lemma 4. As shown in the figure, the simulations perfectly matches the analytical results, thus the validity of the analysis based on the assumption that $l>>R_j$ is verified. It is also evident, as shown in the figure, as the interference power increases, the outage performance of the near users becomes worse, which is consistent with our intuition.}
\begin{figure}[!t]
\setlength{\belowcaptionskip}{-1em}   

\centering
\includegraphics[width=3.5in]{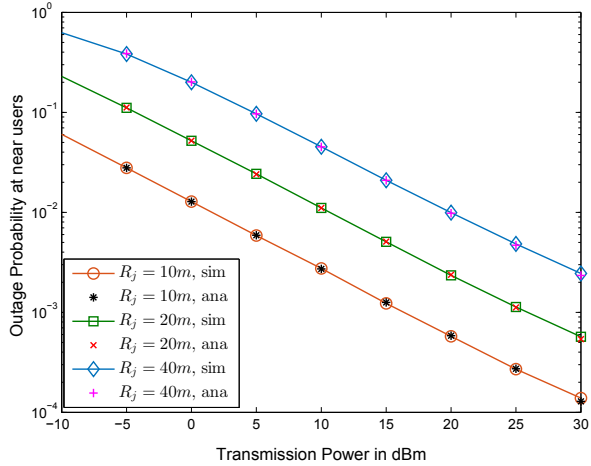}
\caption{Impact of the distance between a BS and its near user on the outage probability at the near user. $1\leq j\leq3$.}
\label{impact_Rj}
\end{figure}

In Fig. \ref{impact_Rj}, the impact of the distance between a BS and its near user on the outage probability achieved by the near user is investigated.
{\color{blue}The parameters are set as: $l=400$m, $R_0=250$m, $r_j=0.5$ BPCU, $0\leq j \leq3$, $P_I=6$dBm, $\lambda_I=\frac{1}{\pi200^2}$, $\beta_0^2=\frac{4}{5}$ and $\alpha=3$.}
As {\color{black} it also concluded from this} figure, given a fixed transmission power, the outage probability achieved by the near user is increased {\color{black}after an increase of } the distance between the BS and its near user. This is reasonable, {\color{black}because} as the distance between a BS and its near user increases, the large scale propagation losses become severe.

\begin{figure*}[!t]
\setlength{\belowcaptionskip}{-1em}   
\centering
\subfloat[User 0]{\includegraphics[width=3.2in]{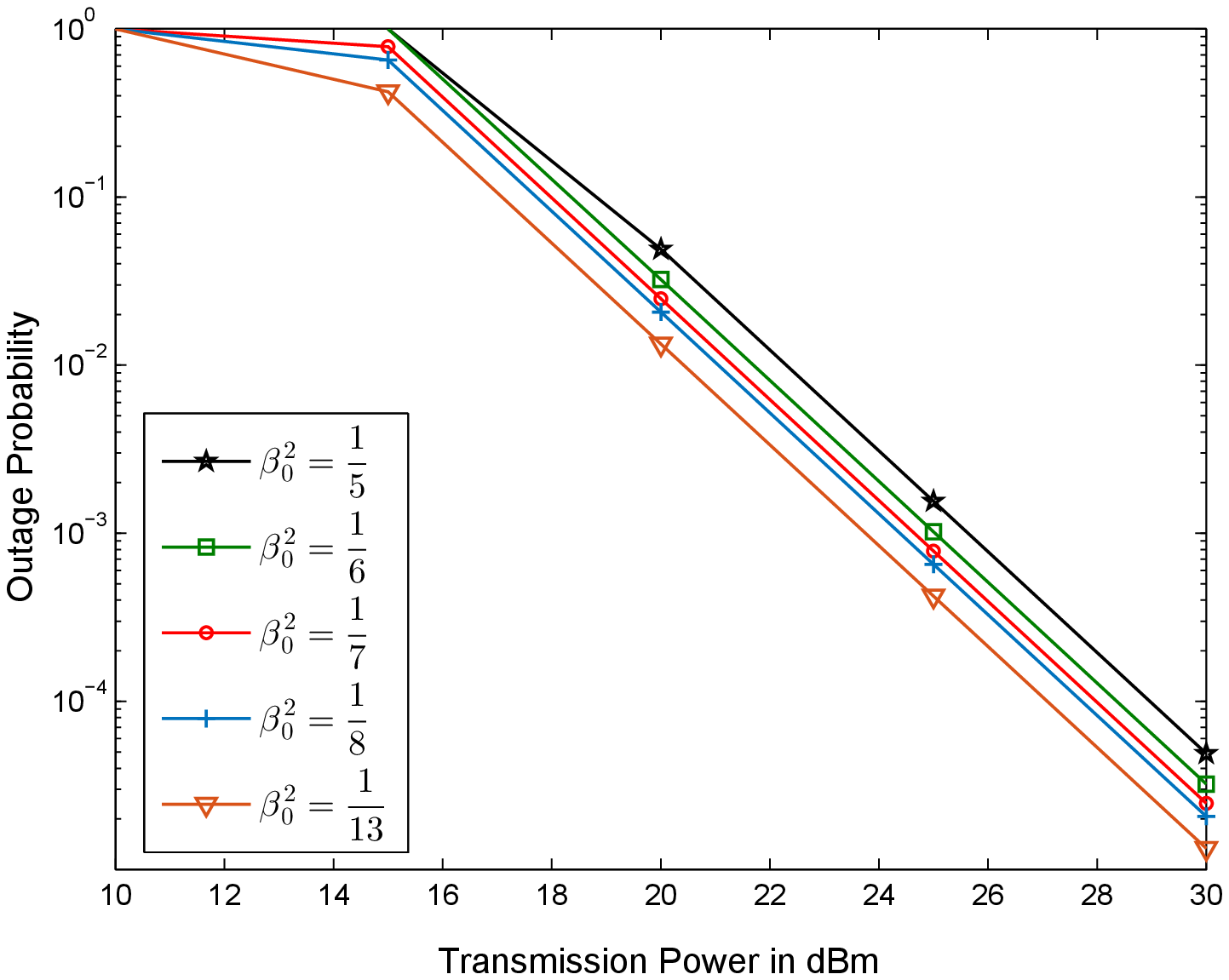}%
\label{beta_a}}
\hfil
\subfloat[User 1,2 {\color{black}and} 3]{\includegraphics[width=3.2in]{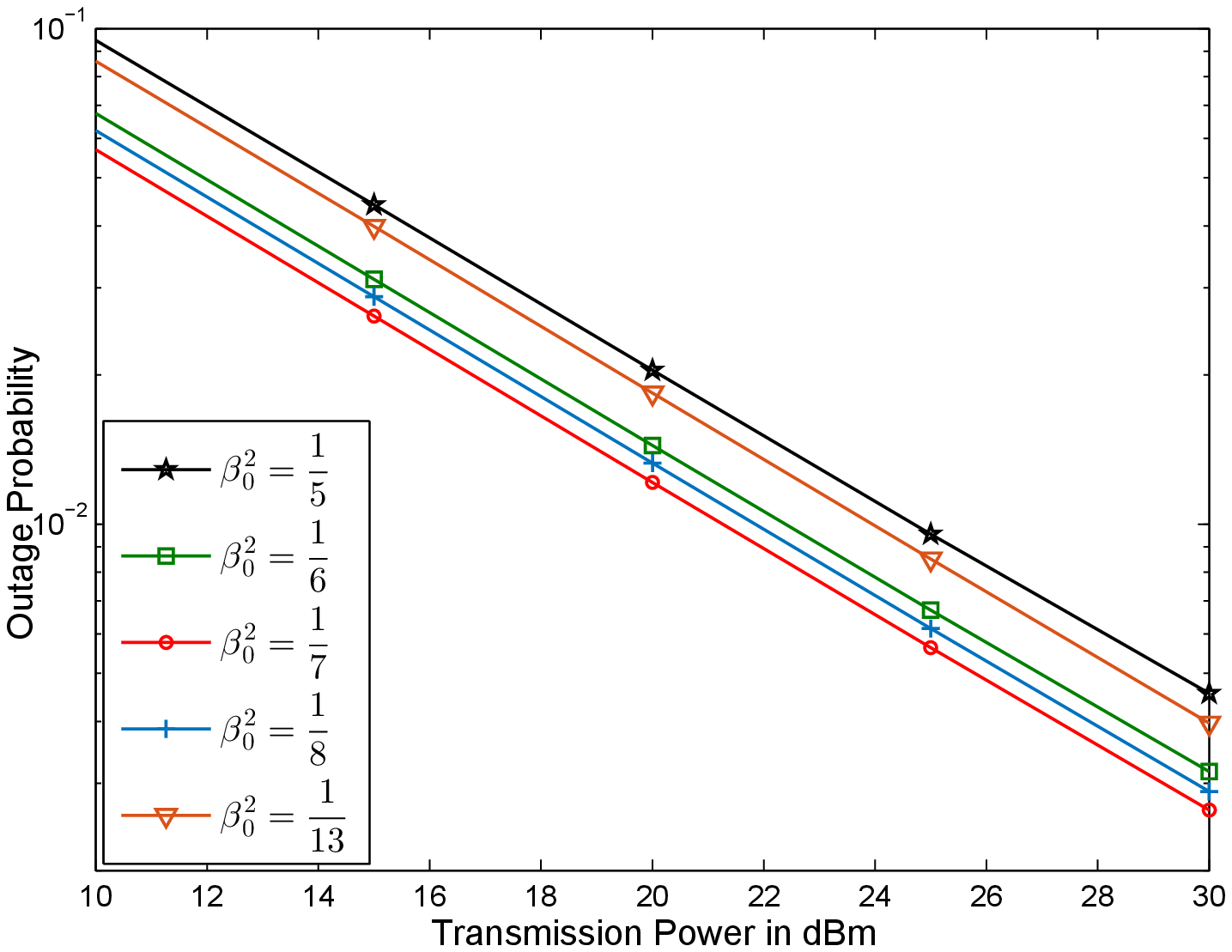}%
\label{beta_b}}
\caption{Impact of power allocation coefficients on the outage probability.}
\label{beta}
\end{figure*}

In Fig. \ref{beta}, the impact of the power allocation coefficients on the outage {\color{black}probability} is studied.
{\color{blue}The parameters are set as: $l=600m$, $R_0=400m$, $R_j=30m, 1\leq j\leq3$,
$r_0=2$ BPCU, $r_1=r_2=r_3=1$ BPCU, $P_I=6$dBm, $\lambda_I=\frac{1}{\pi200^2}$ and $\alpha=3$.}
As it can be seen from Fig. \ref{beta}(a), the outage probability achieved by user 0 increases with $\beta_1$.  Fig. \ref{beta}(b) shows that, when $\beta_1^2<\frac{\eta_j}{\eta_0+\eta_j+\eta_0\eta_j}=\frac{1}{7}$, the outage probabilities achieved by user 1, 2 and 3 decrease with $\beta_1$. On the other hand, when $\beta_1^2>\frac{1}{7}$, the outage probabilities achieved by user 1, 2 and 3 increase with $\beta_1$. Thus the simulation results comfirm the conclusions {\color{black}presented} in  Proposition 2 and Corollary 2.

\section{Conclusion}
In this paper, in order to show the feasibility of N-NOMA, we have proposed a distributed analog beamforming based N-NOMA scheme for a downlink CoMP system with randomly deployed users. Closed-form {\color{black}analysis} of the outage {\color{black}probability}, achieved by the proposed transmission scheme,  {\color{black}has} been developed to facilitate the performance evaluation. The impact of system parameters, such as user locations, distances between BSs and power allocation coefficients, on the outage performance has also been captured.  Computer simulation results have been provided to demonstrate the accuracy of the developed analytical results. The proposed N-NOMA outperforms conventional OMA scheme, as demonstrated by the {\color{black}presented} analytical and computer simulation results. Note that, fixed power allocation coefficients are used in this paper, and it is important to study more sophisticated power allocation strategies {\color{black}in order to} improve the performance of N-NOMA in the future.

{\color{blue}For more general system models, stochastic geometry can be applied. As in \cite{Ding2017NOMAcaching}, A poisson cluster point process(PCP) can be applied, where locations of the base stations are modelled as a homogeneous Poisson point process, each base station is a parent node of a cluster covering a disk, there are K near users randomly distributed in
each cluster. The location of the cell-edge user can be treated as a point of the poisson point process or
can be set at the origin. Applying the proposed N-NOMA to the  stochastic geometry based model will be a potential future direction.}

\appendices
\section{Proof for Lemma1}
{\color{black}In order to evaluate} the outage probability $P_0$, we first fix the location of user 0, i.e., $p_0$. Given the fixed $p_0$, the distribution function of $h_{i0}$ can be obtained. By using the property of the inequality in (\ref{P01}), we get the conditioned outage probability, and then finally we {\color{black}remove the condition} on $p_0$.
\subsection{Characterizing the Outage Probability with a Fixed $p_0$}
Note that, as long as $p_0$ is fixed, $L_{i0}$, $1\leq i\leq3$, can be determined by $p_0$, and each $|h_{i0}|$, $1\leq i\leq3$, is Rayleigh distributed and independent to each other. The conditional pdf of each $|h_{i0}|$, given $p_0$,  can be expressed as
\begin{align}
 f_{|h_{i0}|}(x\left|p_0\right.)=2L_{i0}xe^{-L_{i0}x^2}, x>0.
\end{align}
The joint conditional pdf of $|h_{10}|$, $|h_{20}|$ and $|h_{30}|$ can be expressed as
\begin{align}
  f_{|h_{i0},|h_{20},|h_{30}|}(x,y,z\left|p_0\right.)=8L_{10}L_{20}L_{30}xyze^{-\left(L_{10}x^2+L_{20}y^2+L_{30}z^2\right)}.
\end{align}
Then the outage probability of user 0 can be {\color{black}formulated as}
\begin{align}
  P_0&= E_{p_0}\left\{ \iiint\limits_{\quad(x,y,z)\in V} f_{|h_{i0},|h_{20},|h_{30}|}(x,y,z\left|p_0\right.)\,dx\,dy\,dz\right\} \notag\\
  &= E_{p_0}\left\{ \iiint\limits_{\quad(x,y,z)\in V} 8L_{10}L_{20}L_{30}xyze^{-\left(L_{10}x^2+L_{20}y^2+L_{30}z^2\right)}\,dx\,dy\,dz\right\},
\end{align}
where $V=\left\{ (x,y,z)\left|
 \left( \beta_0^2-\beta_1^2\eta_0\right)\left(x^2+y^2+z^2\right)
                 +2\beta_0^2\left(xy+yz+xz\right)
                 <\frac{\eta_0}{\rho},x>0,y>0,z>0\right. \right\} $
is the integral region. At high SNR, $P_0$ can be approximated as
\begin{align}\label{P_0_app}
P_0&\approx E_{p_0}{\left\{ \iiint\limits_{\quad(x,y,z)\in V} 8L_{10}L_{20}L_{30}xyz\,dx\,dy\,dz\right\}}\notag\\ &=8E_{p_0}\left\{L_{10}L_{20}L_{30}\right\}\iiint\limits_{\quad(x,y,z)\in V} xyz\,dx\,dy\,dz.
\end{align}
Note that, in this paper, as mentioned in Section III, we have assumed that $\beta_0^2-\beta_1^2\eta_0>0$. {\color{black}The integral in (\ref{P_0_app}) can be evaluated by treating first the integral with respect to $x$ and $y$, $z $ as constants. Then the integral can be written as}
\begin{align}\label{Inyz}
\iiint\limits_{\quad(x,y,z)\in V} xyz\,dx\,dy\,dz &=\underset{(y,z)\in V_{yz}}{\iint}\int_{0}^{X(y,z)}xyz\,dx\,dy\,dz \notag\\
&=\frac{1}{2}\underset{Q_1}{\underbrace{\underset{(y,z)\in V_{yz}}{\iint}X^2(y,z)yz\,dy\,dz}},
\end{align}
where $X(y,z)$ and $V_{yz}$ are obtained from $V$ by solving a quadratic inequality problem. {\color{black} The} $X(y,z)$ can be expressed as
\begin{align}
&X(y,z)=\frac{\sqrt{(2\beta_0^2\beta_1^2\eta_0-\beta_1^4\eta_0^2)(y^2+z^2)+2\beta_0^2\beta_1^2\eta_0yz+(\beta_0^2-\beta_1^2\eta_0)\eta_0/\rho}
-\beta_0^2(y+z)}{\beta_0^2-\beta_1^2\eta_0},
\end{align}
{\color{black}and $V_{yz}$ as}
\begin{align}
&V_{yz}=\left\{ (y,z)\left|
 \left( \beta_0^2-\beta_1^2\eta_0\right)\left(y^2+z^2\right)
                 +2\beta_0^2yz
                 <\frac{\eta_0}{\rho},y>0,z>0\right. \right\}.
\end{align}
{\color{black}In order to} simplify the notation of range of the integral, we use the following variable substitution
\begin{align}
\begin{cases}
y=-\frac{\sqrt{2}}{2}u+\frac{\sqrt{2}}{2}v\\
z=\frac{\sqrt{2}}{2}u+\frac{\sqrt{2}}{2}v
\end{cases}.
\end{align}
Now the range of the integral, denoted by $V_{uv}$, can be  easily
derived from $V_{yz}$ as
\begin{align}
V_{uv}&=\left\{ (u,v)\left|
 \left( 2\beta_0^2-\beta_1^2\eta_0\right)v^2
                 -2\beta_1^2\eta_0u^2
                 <\frac{\eta_0}{\rho},v>|u|\right. \right\}.
\end{align}
Then the integral in (\ref{Inyz}) can be {\color{black}written as}
\begin{align}
Q_1&=\frac{1}{2}\underset{(u,v)\in V_{uv}}{\iint}X^2(u,v)(v^2-u^2)\,du\,dv,
\end{align}
where $X(u,v)$ is derived from $X(y,z)$ and
\begin{align}
&X(u,v)=\frac{\sqrt{(\beta_0^2\beta_1^2\eta_0-\beta_1^4\eta_0^2)u^2+(3\beta_0^2\beta_1^2\eta_0-\beta_1^4\eta_0^2)v^2+(\beta_0^2-\beta_1^2\eta_0)\eta_0/\rho}
-\sqrt{2}\beta_0^2v}{\beta_0^2-\beta_1^2\eta_0}.
\end{align}
By further observation, the range of the integral {\color{black}can be} decomposed into two parts:
\begin{itemize}
  \item when $0<v<\sqrt{\frac{\eta_0}{\rho\left( 2\beta_0^2-\beta_1^2\eta_0\right)}}$, the range of $u$ is $-v<u<v$;
  \item when $\sqrt{\frac{\eta_0}{\rho\left( 2\beta_0^2-\beta_1^2\eta_0\right)}}<v<\sqrt{\frac{\eta_0}{\rho\left( 2\beta_0^2-2\beta_1^2\eta_0\right)}}$, the range of $u$ is $\sqrt{\frac{\left( 2\beta_0^2-\beta_1^2\eta_0\right)v^2-\eta_0/\rho}{\beta_1^2\eta_0}}<|u|<v$.
\end{itemize}
Then $Q_1$ can be {\color{black}evaluated} as
\begin{align}
Q_1=\underset{Q_{11}}{\underbrace{\int_{0}^{\phi}\int_{0}^{v}X^2(u,v)(v^2-u^2)\,du\,dv}}
+\underset{Q_{12}}{\underbrace{\int_{\phi}^{\phi_1}\int_{\varphi(v)}^{v}X^2(u,v)(v^2-u^2)\,du\,dv}},
\end{align}
where
\begin{align}
\begin{cases}
\phi_1=\sqrt{\frac{\eta_0}{\rho\left( 2\beta_0^2-2\beta_1^2\eta_0\right)}}\\
\varphi(v)=\sqrt{\frac{\left( 2\beta_0^2-\beta_1^2\eta_0\right)v^2-\eta_0/\rho}{\beta_1^2\eta_0}}\notag
\end{cases}.
\end{align}
An interesting observation is that the second part integral is much smaller than the first part,  i.e., $Q_{11}>>Q_{12}$. Thus $Q_1$ can be approximated as
\begin{align}
Q_1\approx&\int_{0}^{\phi}\int_{0}^{v}X^2(u,v)(v^2-u^2)\,du\,dv.
\end{align}
After some manipulations, {\color{black}we get}
\begin{align}
Q_1\approx\frac{1}{(\beta_0^2-\beta_1^2\eta_0)^2}\left(G(\phi)-F(\phi)+F(0)\right).
\end{align}
Then, the outage probability at user 0 can be expressed as
\begin{align}
P_0&=4Q_1E_{p_0}\left\{L_{10}L_{20}L_{30}\right\}.
\end{align}
\subsection{Removing the condition on $p_0$}
The next step of the proof is to remove the condition on the location of user 0. To accomplish this, recall that user 0 is uniformly distributed in the intersecting area of the three discs, i.e., $A$. Therefore, the above expectation with respect to $p_0$ can be {\color{black}written as}
\begin{align}
E_{p_0}\left\{L_{10}L_{20}L_{30}\right\}&=\int_{p_0 \in A}L_{10}L_{20}L_{30}\frac{1}{S_A}\,d{p_0}\notag\\
           &=\frac{1}{S_A}\int_{p_0 \in A}L_{10}L_{20}L_{30}\,d{p_0},
\end{align}
where $S_A$ is the area of A.
Furthermore, recall that $A$ is composed of $A_1$, $A_2$ and $A_3$ as illustrated in Fig. \ref{system}. Due to the symmetry of $A_1$, $A_2$ and $A_3$, the expectation can be further {\color{black}formulated} as
\begin{align}
E_{p_0}\left\{L_{10}L_{20}L_{30}\right\}=&\frac{1}{3}\left(
              \frac{1}{S_{A_1}}\int_{p_0 \in A_1}L_{10}L_{20}L_{30}\,d{p_0}+\right.\notag\\
              &\left.\frac{1}{S_{A_2}}\int_{p_0 \in A_2}L_{10}L_{20}L_{30}\,d{p_0}+
              \frac{1}{S_{A_3}}\int_{p_0 \in A_3}L_{10}L_{20}L_{30}\,d{p_0} \right)\notag\\
           =&\frac{1}{S_{A_1}}\int_{p_0 \in A_1}L_{10}L_{20}L_{30}\,d{p_0},
\end{align}
where $S_{A_1}$ is the area of $A_1$, {\color{black}given by}
\begin{align}
   S_{A_1}=\frac{1}{3}S_A=R_0^2\left(\frac{\pi}{3}-\arcsin\left(\frac{l}{2R_0}\right)\right)-\frac{\sqrt{3}}{3}lR_0\sin{\left(\frac{\pi}{3}-\arcsin\left(\frac{l}{2R_0}\right)\right)}.
\end{align}
\begin{figure}[!t]
\setlength{\belowcaptionskip}{-1em}   
\centering
\includegraphics[width=2.5in]{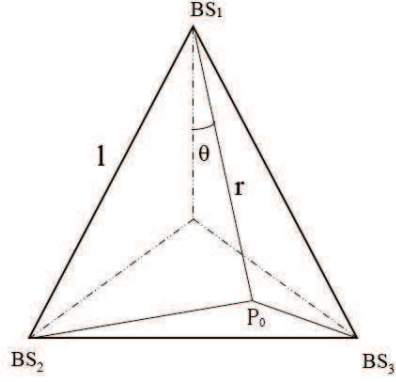}
\caption{{\color{black}Distances between BSs and user 0}}
\label{polar_1}
\end{figure}After {\color{black}transforming} to polar coordinates, as shown in Fig. \ref{polar_1}, $L_{10}$, $L_{20}$ and $L_{30}$ can be expressed as
\begin{align}
\begin{cases}
L_{10}=r^\alpha\\
L_{20}=(l^2+r^2-2lrcos(\pi/6+\theta))^{\frac{\alpha}{2}}\\
L_{30}=(l^2+r^2-2lrcos(\pi/6-\theta))^{\frac{\alpha}{2}}
\end{cases},
\end{align}
which are obtained by using {\color{black}the law of cosines}. Then, the expectation can be {\color{black}written as}
\begin{align}\label{E_L3}
E_{p_0}\left\{L_{10}L_{20}L_{30}\right\}&=\frac{1}{S_{A_1}}\int_{\frac{\sqrt{3}}{3}l}^{R_0}\int_{-\Theta}^{+\Theta}L_{10}L_{20}L_{30}r\,d\theta\,dr,
\end{align}
where $\Theta$ {\color{black}is}
\begin{align}
  \Theta=\frac{\pi}{3}-\arcsin\left(\frac{l}{2r}\right).
\end{align}
{\color{blue}After some algebraic manipulations, the expectation expression in (\ref{E_L}) for $\alpha=2$ can be obtained. For the case when $\alpha>2$, we only focus on the scenario when $\frac{R_0}{l} \to\frac{\sqrt{3}}{3}$, which yields $\theta\to0$. By using Taylor series, the product of $L_{10}$, $L_{20}$ and $L_{30}$ can be expressed as
\begin{align}
L_{10}L_{20}L_{30}=&\underset{\tilde{Q}_1}{\underbrace{\frac{1}{2} \alpha \left(\sqrt{3}
   l^3 r^3-4 l^2 r^4+\sqrt{3}
   l r^5\right)+\left(l^4
   r^2-2 \sqrt{3} l^3 r^3+5
   l^2 r^4-2 \sqrt{3} l
   r^5+r^6\right)^{\alpha/2}}}\notag\\
   &+O(\theta^2),
\end{align}
thus, (\ref{E_L3}) can be further calculated as
\begin{align}\label{E_L4}
E_{p_0}\left\{L_{10}L_{20}L_{30}\right\}\approx
\frac{1}{S_{A_1}}\int_{\frac{\sqrt{3}}{3}l}^{R_0}2\Theta \tilde{Q}_1r\,dr,
\end{align}
Replacing r with x in (\ref{E_L4}), where $r=(\frac{\sqrt{3}}{3}+x)l$. Note that when $\frac{R_0}{l}\to\frac{\sqrt{3}}{3}$, $x\to0$. Thus we get the following approximation:
\begin{align}
2\Theta\tilde{Q}_1r\approx 2\times3^{\frac{1}{2}-\frac{3\alpha}{2}}l^{3\alpha+1}x
\end{align}
which is obtained by using Taylor series. Then $E_{p_0}\left\{L_{10}L_{20}L_{30}\right\}$ can be calculated as
\begin{align}
E_{p_0}\left\{L_{10}L_{20}L_{30}\right\}&\approx
\frac{1}{S_{A_1}}\int_{0}^{\frac{R_0}{l}-\frac{\sqrt{3}}{3}}2\times3^{\frac{1}{2}-\frac{3\alpha}{2}}l^{3\alpha+2}x\,dx,\notag\\
&=\frac{\sqrt{3}(R_0-\frac{\sqrt{3}}{3}l)^2\times3^{-\frac{3\alpha}{2}}l^{3\alpha}}{S_{A_1}}\notag\\
&\overset{(a)}{\approx}3^{-\frac{3\alpha}{2}}l^{3\alpha}
\end{align}
where (a) follows from the fact that $S_{A_1}$ can be approximated as $S_{A_1}\approx\sqrt{3}\left(R_0-\frac{\sqrt{3}}{3}l\right)^2$, when $\frac{R_0}{l}\to\sqrt{3}/3$. Therefore Lemma 1 is proved.}
\section{proof for Lemma 2}
The received signal can be expressed as
\begin{align}
\tilde{y_0}=\sqrt{\sum\limits_{i=1}^{3}{|h_{i0}|^2}}\sqrt{3P_s}s_0+n_0
\end{align}
{\color{black}and} the received SNR {\color{black}as}
\begin{align}
\text{SNR}=\frac{3P_s\sum\limits_{i=1}^{3}{|h_{i0}|^2}}{\sigma^2}
 =3\rho\sum\limits_{i=1}^{3}{|h_{i0}|^2}.
\end{align}
Then, the outage probability is given by
\begin{align}
\tilde{P_0}=P\left( \log\left(1+3\rho\sum\limits_{i=1}^{3}{|h_{i0}|^2}\right) < r_0 \right)
           =P\left(  \sum\limits_{i=1}^{3}{|h_{i0}|^2} < \frac{\eta_0}{3\rho}        \right).
\end{align}
The key step to calculate $\tilde{P_0}$ is to obtain the pdf of $\sum_{i=1}^{3}|h_{i0}|^2$.
Note that, as long as the location of user 0 ($p_0$) is fixed, $L_{i0}$, $1 \leq i \leq 3$, can be determined by $p_0$,  and the pdf of $|h_{i0}|^2$ given $p_0$ {\color{black}follows} exponential distribution {\color{black}with pdf given by}
\begin{align}
f_{|h_{i0}^2|}(x|p_0)=L_{i0}e^{-L_{i0}x}.
\end{align}
Define $\psi=\sum_{i=1}^{3}|h_{i0}|^2$. Note that if $L_{10}\neq L_{20}\neq L_{30}$, then the pdf of $\psi$ is given by
\begin{align}
f_{\psi}(x|p_0) = \sum\limits_{i=1}^{3}{ L_{i0}e^{-L_{i0}x}  \prod\limits_{\substack{j=1\\j\neq i}}^{3}{\frac{L_{j0}}{L_{j0}-L_{i0}}} }.
\end{align}
Thus the outage probability can be calculated as follows:
\begin{align}
\tilde{P_0}=E_{p_0}\left\{ \int_{0}^{\frac{\eta_0}{3\rho}} f_{\psi}(x|p_0)\,dx\right\}
=   E_{p_0}\left\{\underset{Q_3}{ \underbrace{\sum\limits_{i=1}^{3}{ \left(1-e^{-L_{i0}\frac{\eta_0}{3\rho}}\right)  \prod\limits_{\substack{j=1\\j\neq i}}^{3}{\frac{L_{j0}}{L_{j0}-L_{i0}}} }}} \right\}.
\end{align}
By changing to polar coordinates, the above expectation with respect to $p_0$ can be {\color{black}evaluated} as
\begin{align}\label{A3_1}
  \tilde{P_0}&=\frac{1}{S_{A_1}}\int_{\frac{\sqrt{3}}{3}l}^{R_0}\int_{-\Theta}^{\Theta}
              Q_3\,rdrd\theta,
\end{align}
where $Q_3$ is a function of $r$ and $\theta$, and $\Theta=\frac{\pi}{3}-\arcsin\left(\frac{l}{2r}\right)$. Note that, we only focus of the scenario of $k\to\sqrt{3}/3 $, which yields
$\theta\to0$. By using {\color{black}the} Taylor series, $Q_3$ can be approximated as{\color{blue}
\begin{align}
Q_3&\approx \frac{r^{\alpha } e^{-t T}
   \left(r^{\alpha } \left(e^{t
   T}-1\right)-T \left(t r^{\alpha }+2
   e^{t T}-2\right)+t T^2\right)+T^2
   \left(1-e^{t \left(-r^{\alpha
   }\right)}\right)}{\left(T-r^{\alpha
   }\right)^2}\notag\\
   & \overset{\Delta}{=}\tilde{Q}_3,
\end{align}
where $T=\left(l^2-\sqrt{3}lr+r^2\right)^{\alpha/2}$. Note that $\tilde{Q}_3$ is a function of only $r$.
Then (\ref{A3_1}) can be further written as
\begin{align}\label{P_0_61}
  \tilde{P_0}\approx\frac{1}{S_{A_1}}\int_{\frac{\sqrt{3}}{3}l}^{R_0}\int_{-\Theta}^{\Theta}
              \tilde{Q_3}\,rdrd\theta
             \approx\frac{1}{S_{A_1}}\int_{\frac{\sqrt{3}}{3}l}^{R_0}
              2\Theta\tilde{Q_3}\,rdr.
\end{align}
Replacing $r$ with $x$ in (\ref{P_0_61}), where $r=\frac{\sqrt{3}}{3}l(1+x)$. Note that when $k\to\sqrt{3}/3 $, $x\to 0$. Thus we get the following approximation:
\begin{align}
 2\Theta\tilde{Q_3}r\approx \frac{ e^{-\frac{3^{-\frac{\alpha }{2}-1} \eta_0
    l^{\alpha }}{\rho }} \left(-\eta_0^2 l^{2 \alpha }+2\times
   3^{\alpha +2} \rho ^2 \left(e^{\frac{3^{-\frac{\alpha
   }{2}-1} \eta_0 l^{\alpha }}{\rho }}-1\right)-2\times
   3^{\frac{\alpha }{2}+1} \eta_0  \rho  l^{\alpha }\right)\sqrt{3}lx}{3^{\alpha+2}
   \rho ^2},
\end{align}
which is obtained by using the Taylor series and omitting terms of higher order. Then $\tilde{p_0}$ can be expressed after some algebraic manipulations as
\begin{align}
\tilde{P_0}&\approx \frac{ e^{-\frac{3^{-\frac{\alpha }{2}-1} \eta_0
    l^{\alpha }}{\rho }} \left(-\eta_0 ^2 l^{2 \alpha }+2\times
   3^{\alpha +2} \rho ^2 \left(e^{\frac{3^{-\frac{\alpha
   }{2}-1} \eta_0  l^{\alpha }}{\rho }}-1\right)-2\times
   3^{\frac{\alpha }{2}+1} \eta_0  \rho  l^{\alpha }\right)\sqrt{3}\left(R_0-\frac{\sqrt{3}}{3}l\right)^2}{2\times3^{\alpha+2}
   \rho ^2S_{A_1}}\notag\\
 &\overset{(a)}{\approx}\frac{ e^{-\frac{3^{-\frac{\alpha }{2}-1} \eta_0
    l^{\alpha }}{\rho }} \left(-\eta_0 ^2 l^{2 \alpha }+2\times
   3^{\alpha +2} \rho ^2 \left(e^{\frac{3^{-\frac{\alpha
   }{2}-1} \eta_0  l^{\alpha }}{\rho }}-1\right)-2\times
   3^{\frac{\alpha }{2}+1} \eta_0  \rho  l^{\alpha }\right)}{2\times3^{\alpha+2}
   \rho ^2}
\end{align}
where (a) follows from the fact that $S_{A_1}$ can be approximated as $S_{A_1}\approx\sqrt{3}\left(R_0-\frac{\sqrt{3}}{3}l\right)^2$, when $k\to\sqrt{3}/3$. Therefore, Lemma 2 is proved.}
\section{Proof for lemma 3}
 Without loss of generality, the following derivation process focus on calculating the outage probability achieved by user 1.

Recall that $\left|\tilde{h}_{jj}\right|^2$ is exponentially distributed, i.e., the pdf of  $\left|\tilde{h}_{jj}\right|^2$ is given by
\begin{align}
f_{\left|\tilde{h}_{jj}\right|^2}(x)=L_{ij}e^{-L_{ij}x}.
\end{align}
Then, the outage probability at user 1 can be calculated as
\begin{align}\label{P_1_67}
P_1=E_{p_1,|\tilde{h_{21}}|^2,|\tilde{h_{31}}|^2}
\left\{1-e^{-L_{11}M_1\left[\left(|\tilde{h_{21}}|^2+|\tilde{h_{31}}|^2\right)\beta_1^2+1/\rho\right]}\right\}.
\end{align}
By using (\ref{P_1_67}), the outage probability at user 1 can be expressed as
\begin{align}
P_1=1-E_{p_1}
\left\{e^{-\frac{L_{11}M_1}{\rho}} \times \underset{Q_2}{\underbrace{\frac{L_{21}}{L_{11}M_1\beta_1^2+L_{21}}
\times \frac{L_{31}}{L_{11}M_1\beta_1^2+L_{31}}}} \right\}.
\end{align}
Then, the expectation part can be {\color{black}evaluated} as
\begin{align}
E_{p_1}\left\{Q_2e^{-\frac{L_{11}M_1}{\rho}}\right\}&=\frac{1}{\pi R_1^2}\int_{p_1 \in D_1}Q_2e^{-\frac{L_{11}M_1}{\rho}}\,dp_1\notag\\
&{\overset{(a)}{=}}\frac{1}{\pi R_1^2}\int_{0}^{2\pi}\int_{0}^{R_1}Q_2e^{-\frac{L_{11}M_1}{\rho}}r\,\,drd\theta,
\end{align}
\begin{figure}[!t]
\setlength{\belowcaptionskip}{-1em}   
\centering
\includegraphics[width=2.5in]{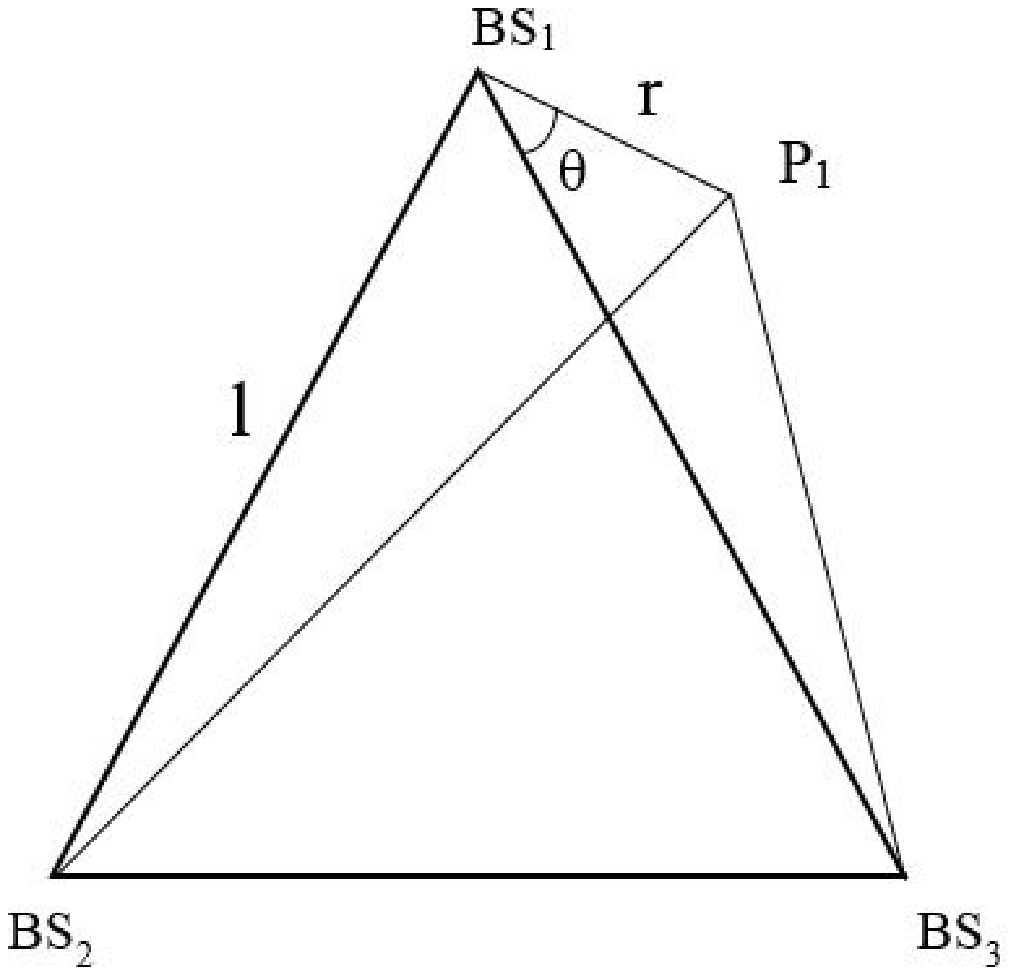}
\caption{{\color{black}Distances between BSs and user 1}}
\label{polar_2}
\end{figure}where (a) follows a step changing to polar coordinates, as shown in Fig. \ref{polar_2}. Note that $L_{11}$, $L_{21}$ and $L_{31}$ can be expressed as{\color{blue}
\begin{align}
\begin{cases}
L_{11}=r^\alpha\\
L_{21}=\left(r^2+l^2-2lr\cos{(\pi/3+\theta)}\right)^\frac{\alpha}{2}\\
L_{31}=\left(r^2+l^2-2lr\cos{(\theta)}\right)^\frac{\alpha}{2}
\end{cases}.
\end{align}
Then, $Q_2$ can be formulated as
\begin{align}
Q_2=&\left(1-\frac{M_1\beta_1^2L_{11}}{M_1\beta_1^2L_{11}+L_{21}}\right)\times\left(1-\frac{M_1\beta_1^2L_{11}}{M_1\beta_1^2L_{11}+L_{31}}\right)\notag\\
   =&\left(1-\frac{M_1\beta_1^2r^\alpha}{M_1\beta_1^2r^\alpha+\left(r^2+l^2-2lr\cos{(\pi/3+\theta)}\right)^\frac{\alpha}{2}}\right)\notag\\
   &\times\left(1-\frac{M_1\beta_1^2r^\alpha}{M_1\beta_1^2r^\alpha+\left(r^2+l^2-2lr\cos{(\theta)}\right)^\frac{\alpha}{2}}\right)
\end{align}
When $R_2<<l$, $Q_2$ can be approximated as
\begin{align}\label{Q_2}
Q_2\approx 1-2\beta_1^2M_1\frac{r^\alpha}{l^\alpha},
\end{align}
which is obtained by {\color{blue}the} using Taylor series. Then the above expectation can be {\color{blue}written} as
\begin{align}
E_{p_1}\left\{Q_2e^{-\frac{L_{11}M_1}{\rho}}\right\}&\approx
\frac{1}{\pi R_1^2}\int_{0}^{2\pi}\int_{0}^{R_1}\left(1-2\beta_1^2M_1\frac{r^\alpha}{l^\alpha}\right)e^{-\frac{r^\alpha M_1}{\rho}}r\,\,drd\theta\\\notag
&=\frac{2\rho^\frac{2}{\alpha}M_1\gamma(\frac{2}{\alpha},\frac{M_1R_1^\alpha}{\rho})
  -\frac{4\beta_1^2M_1\rho^{\frac{2}{\alpha}+1}}{l^\alpha}\gamma(\frac{2}{\alpha}+1,\frac{M_1R_1^\alpha}{\rho})}{\alpha M_1^{\frac{2}{\alpha}+1}R_1^2}.
\end{align}
So, the outage probability at user 1 can be expressed as
\begin{align}
P_1\approx1-\frac{2\rho^\frac{2}{\alpha}M_1\gamma(\frac{2}{\alpha},\frac{M_1R_1^\alpha}{\rho})
  -\frac{4\beta_1^2M_1\rho^{\frac{2}{\alpha}+1}}{l^\alpha}\gamma(\frac{2}{\alpha}+1,\frac{M_1R_1^\alpha}{\rho})}{\alpha M_1^{\frac{2}{\alpha}+1}R_1^2},
\end{align}}
By replacing the subscript of 1 with j, Lemma 3 is proved.
{\color{blue}\section{Proof for lemma 4}
As in Appendix C, this appendix focuses on calculating the outage probability achieved by user 1.

Following the similar steps as in Appendix C, $P_1^{\text{Inter}}$ can be expressed as
\begin{align}\label{P_1_appendix}
P_1^{\text{Inter}}=1-E_{p_1}
\left\{e^{-\frac{L_{11}M_1}{\rho}} \times \underset{Q_2}{\underbrace{\frac{L_{21}}{L_{11}M_1\beta_1^2+L_{21}}
\times \frac{L_{31}}{L_{11}M_1\beta_1^2+L_{31}}}}
\times E_{I_1}\left\{e^{-L_{11}M_1I_1}\right\}\right\}.
\end{align}
Note that, $E_{I_1}\left\{e^{-L_{11}M_1I_1}\right\}$ can be evaluated as
\begin{align}
E_{I_1}\left\{e^{-L_{11}M_1I_1}\right\}&=E_{\Phi_I,g_{I_k,1}}\left\{
\prod\limits_{p_{I_k}\in\Phi_I}\exp\left(-L_{11}M_1\rho_I\frac{|g_{I_k,j}|^2}{L(||p_{I_k}-p_1||)}\right)\right\}\\\notag
&=E_{\Phi_I}\left\{
\prod\limits_{p_{I_k}\in\Phi_I}\frac{1}{\frac{L_{11}M_1\rho_I}{L(||p_{I_k}-p_1||)}+1}\right\}.
\end{align}
By applying Campell theorem and the PFGL, $E_{I_1}\left\{e^{-L_{11}M_1I_1}\right\}$ can be calculated as
\begin{align}
E_{I_1}\left\{e^{-L_{11}M_1I_1}\right\}&=\exp\left(-\lambda_I\int_{R^2}\left(1-\frac{1}{\frac{L_{11}M_1\rho_I}{L(||p-p_1||)}+1}\right)\,dp\right).
\end{align}
By doing the substitution $p'=p-p_1$, we have
\begin{align}
E_{I_1}\left\{e^{-L_{11}M_1I_1}\right\}&=\exp\left(-\lambda_I\int_{R^2}\left(1-\frac{1}{\frac{L_{11}M_1\rho_I}{L(||p'||)}+1}\right)\,dp'\right)\\\notag
&=\exp\left(-\lambda_I\int_{0}^{2\pi}\int_{0}^{\infty}\left(1-\frac{1}{\frac{L_{11}M_1\rho_I}{r^\alpha}+1}\right)r\,drd\theta\right)\\\notag
&=\exp\left(-2\pi\lambda_I\int_{0}^{\infty}\frac{L_{11}M_1\rho_Ir}{L_{11}M_1\rho_I+r^\alpha}\,dr\right)\\\notag
&=\exp\left(-2\pi\lambda_I\frac{(L_{11}M_1\rho_I)^{\frac{2}{\alpha}}}{\alpha}B\left(\frac{2}{\alpha},1-\frac{2}{\alpha}\right)\right),
\end{align}
where the last step is obtained by applying Beta function. Again changing to polar coordinates and using the the approximation as in (\ref{Q_2}), $P_1^{\text{Inter}}$ can be approximated as
\begin{align}
P_1^{\text{Inter}}&\approx1-\frac{1}{\pi R_1^2}\int_{0}^{2\pi}\int_{0}^{R_1}\left(1-2\beta_1^2M_1\frac{r^\alpha}{l^\alpha}\right)\times\\\notag
&\exp\left(-\frac{M_1}{\rho}r^\alpha
-2\pi\lambda_I\frac{(M_1\rho_I)^{\frac{2}{\alpha}}}{\alpha}B\left(\frac{2}{\alpha},1-\frac{2}{\alpha}\right)r^2\right)r\,\,drd\theta.
\end{align}
By Applying Gauss-Chebyshev quadrature and substituting the subscript $1$ with $j$, Lemma 4 follows.}
\bibliographystyle{IEEEtran}
\bibliography{IEEEabrv,ref}
\end{document}